\pdfoutput=1
\documentclass[aps,prl,twocolumn,groupeaddress,nofootinbib,10pt]{revtex4-2}
\usepackage{array}
\usepackage{titlesec}
\titleformat{\paragraph}
{\normalfont\normalsize\bfseries}{\theparagraph}{1em}{}
\titlespacing*{\paragraph}
{0pt}{3.25ex plus 1ex minus .2ex}{1.5ex plus .2ex}
\renewcommand*{\theparagraph}{\roman{paragraph})}

\usepackage{tikz-feynman}

\usepackage{amssymb, amsmath,amsthm,bm}    
\usepackage{graphicx}   
\usepackage{verbatim}   
\usepackage{color}      
\usepackage[labelformat=simple]{subcaption}
\usepackage{mathrsfs} 
\usepackage[linktocpage = true]{hyperref}  
\usepackage{cancel}
\usepackage{graphicx}
\usepackage{caption} 
\usepackage{tikz}
\usepackage{tcolorbox}
\usepackage{pgflibraryarrows}
\usepackage{pgflibrarysnakes}
\usepackage{yfonts}
\usepackage{multirow}
\usetikzlibrary{decorations.markings}
\usetikzlibrary{decorations.footprints}
\definecolor{green}{RGB}{35,142,35}
\usetikzlibrary{decorations.pathmorphing}
\usepackage{float}
\usepackage{appendix}

\usepackage{empheq} 
 
\usepackage{ulem}

\usepackage[framemethod=tikz]{mdframed}

\hypersetup{
    colorlinks   = true,
    linkcolor = blue,
    citecolor    = red
}  
\makeatletter
\def\p@subsection{}
\makeatother

\newcommand{\etaphi}{\eta^{(\Phi)}}
\newcommand{\tJV}{\tilde{J}^V}
\newcommand{\tgf}{\tilde{g}_{f}}
\newcommand{\tgfi}{\tilde{g}_{f,i}}
\newcommand{\sez}{\sqrt{\ell_0}}

\usepackage{cleveref}

\allowdisplaybreaks

\label{eq:partition_function_general2}

\newtcolorbox{mymathbox}[1][]{colback=white, #1}

\usepackage[framemethod=tikz]{mdframed}

\newcommand{\ncmd}{\newcommand}
\ncmd{\lt}{\left}
\ncmd{\rt}{\right}
\newcommand{\eq}[1]{Eq. \eqref{#1}}
\newcommand{\fig}[1]{Fig. \ref{#1}}
\newcommand{\tr}[1]{\mbox{Tr}\lt[{#1}\rt]}

\ncmd{\kF}{$k_F$ }
\ncmd{\Lf}{$\Lambda_f$ }
\ncmd{\Lb}{$\Lambda_b$ }
\ncmd{\KF}{k_{\mathrm{F}}}

\definecolor{darkblue}{RGB}{10,10,150}
\DeclareMathOperator{\sgn}{sgn}

\newcommand{\dd}{\mathrm{d}}

\makeatletter
\newcommand*{\rom}[1]{\expandafter\@slowromancap\romannumeral #1@}
\makeatother

\makeatletter
\renewcommand\theequation{{\color{blue} \theequation@prefix \arabic{equation}}}
\makeatother

\makeatletter
\def\p@subsection{}
\makeatother

\newcommand{\bqa}{\begin{eqnarray}} 
\newcommand{\eqa}{\end{eqnarray}}

\newcommand{\spmqty}[1]{\begin{psmallmatrix}#1\end{psmallmatrix}}

\newcommand{\abs}[1]{\left| #1 \right|}

\begin{document}

\title{ 
Ultraviolet/infrared mixing-driven 
suppression of Kondo screening \\
in the antiferromagnetic quantum critical metal 
}
        \author{Francisco Borges$^{1,2}$\footnote{fborges@perimeterinstitute.ca} }
        \author{Peter Lunts$^{3}$}
        \author{Sung-Sik Lee$^{1,2}$\footnote{slee@mcmaster.ca}}
        \affiliation{$^{1}$Department of Physics \& Astronomy, McMaster University, Hamilton ON L8S 4M1, Canada}
        \affiliation{$^{2}$Perimeter Institute for Theoretical Physics, Waterloo ON N2L 2Y5, Canada}
        \affiliation{$^{3}$Department of Physics, Harvard University, Cambridge MA 02138, USA}
        
        \date{May 6, 2025}
        
\begin{abstract}
We study a magnetic impurity immersed in the two-dimensional antiferromagnetic quantum critical metal (AFQCM),
using the field-theoretic functional renormalization group.
Critical spin fluctuations represented by a bosonic field compete with itinerant electrons to couple with the impurity through the spin-spin interaction.
At long distances, the antiferromagnetic electron-impurity (Kondo) coupling dominates over the boson-impurity coupling.
However, the Kondo screening is weakened by the boson 
with an increasing severity
as the hot spots connected by the magnetic ordering wave-vector are better nested.
For $v_{0,i} \ll 1$, where $v_{0,i}$ is the bare nesting angle at the hot spots, 
the temperature 
$T_K^{\mathrm{AFQCM}}$
below which Kondo coupling becomes $O(1)$ is suppressed as $\frac{\log \Lambda/T_K^{\mathrm{AFQCM}}}{\log \Lambda/T_K^{\mathrm{FL}}} \sim \frac{g_{f,i}}{v_{0,i} \log 1/v_{0,i} }$,
where $T_K^{\mathrm{FL}}$ is the Kondo temperature of the Fermi liquid with the same electronic density of states,
and $g_{f,i}$ is the boson-impurity coupling defined at UV cutoff energy $\Lambda$.
The remarkable efficiency of the single collective field 
in hampering the screening of the impurity spin 
by the Fermi surface 
originates from 
a ultraviolet/infrared (UV/IR) mixing: 
bosons with momenta up to a UV cutoff actively suppress Kondo screening at low energies.
\end{abstract}

\maketitle
%
%
%




Collective phenomena that arise from the interplay between local magnetic moments and itinerant electrons continue to attract considerable attention in condensed matter physics \cite{
PhysRev.124.41,
10.1143/PTP.32.37,
Nozieres:1974aa,
DONIACH1977231,
Hewson1993,
RevModPhys.56.755,
RevModPhys.59.1,
Si_Steglich_heavy_fermions_review,
PhysRevB.58.3794,
Senthil_Sachdev_Vojta_2003,
Senthil_Vojta_Sachdev_2004,
Coleman_topological_kondo_ins,
Kondo_destruction_Si_Pixley,
Checkelsky_et_al_flat_bands_Kondo_review,
Georges_Kotliar_Hunds_popular,
Coleman_AFM_heavy_fermions,
PhysRevB.88.245111,
PhysRevX.10.041021}. 
In the classic Kondo problem, a magnetic impurity put in Fermi liquid metals is screened into a spin singlet 
due to the quantum effect that is akin to what causes the confinement of quarks \cite{
10.1143/PTP.32.37,
Kondo_problem_RevModPhys}.
Recent discoveries of correlated compounds that can be tuned across quantum phase transitions have opened opportunities for exploring new cooperative magnetism 
facilitated by 
critical fluctuations \cite{
Larkin_Melnikov,
Maebashi_Varma_Ferro_2002,
Varma_Kondo_effect_AFM_2005,
Varma_Kondo_effect_AFM_phase_2008,
PhysRevLett.79.3755,
CUSTERS}.
However, the current theoretical understanding remains limited because quantum critical metals are usually described by strongly interacting theories \cite{SUNGSIKREVIEW}. 

In this paper, we study a single magnetic impurity immersed in the two-dimensional antiferromagnetic quantum critical metal (AFQCM)\cite{
ABANOV3,
MAX2}
based on the recent non-perturbative understanding of the strongly coupled theory\cite{
SCHLIEF,
BORGES2023169221}.
This problem has been studied in Ref. \cite{Varma_Kondo_effect_AFM_2005}, but we find that strong coupling effects 
make a qualitative difference for the behavior of the impurity.
In AFQCM,
itinerant electrons and critical spin fluctuations compete to get entangled with the impurity spin.
One may naively expect that critical fluctuations, described by bosonic fields, is no match for itinerant electrons with an extensive Fermi surface.
However, critical spin fluctuations turn out to be almost on the par with the Fermi surface in the competition.
Through the strong electron-boson coupling,
the boson `recruits' 
low-energy particle-hole excitations with a wide range of momenta 
to enhance its density of states and suppress Kondo screening at low energies.
While Kondo coupling still becomes strong at long distances, it only happens at a much larger distance scale compared to Fermi liquids when the Fermi surface connected by the antiferromagnetic wavevector is close to nesting.

\begin{figure*}[t]
    \centering
    \begin{subfigure}{0.3\linewidth}
    \centering
        \includegraphics[width=\linewidth]{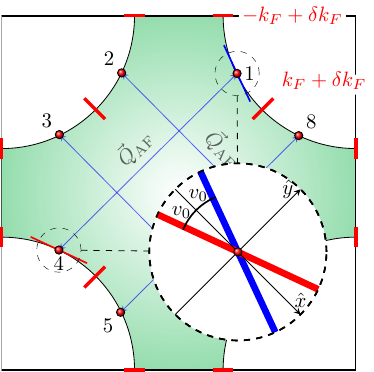}
        \caption{\label{fig:HS}}
    \end{subfigure}
    \begin{subfigure}{0.3\linewidth}
    \centering
        \includegraphics[width=\linewidth]{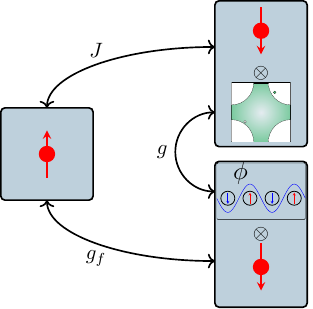}
        \caption{\label{fig:setup}}
    \end{subfigure}
    \begin{subfigure}{0.3\linewidth}
    \centering
        \includegraphics[width=\linewidth]{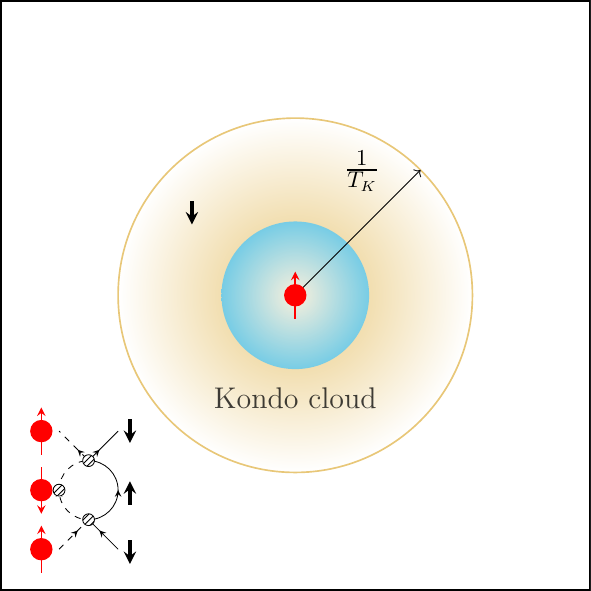}
        \caption{\label{fig:singlet}}
    \end{subfigure}
    \caption{
     ({\color{blue}$a$}) 
In AFQCM, the hot spots (red dots) connected by the antiferromagnetic wave-vector $\vec Q_{AF}$ are strongly coupled with the boson that represents critical spin fluctuations.
$v_0$, which represents the nesting angle between the pairs of hot spots connected by $\vec Q_{AF}$,
determines the low-energy dynamics of the clean AFQCM.
({\color{blue}$b$})
The electron-impurity coupling ($J$) 
and boson-impurity coupling ($g_f$)
cause the impurity spin to flip by creating a particle-hole excitation and a boson, respectively, while
the latter two are strongly mixed through the electron-boson coupling ($g$).
({\color{blue}$c$})
The Kondo temperature vanishes in the small $v_0$ limit due to the dressing of the impurity by the critical spin fluctuations subject to strong UV/IR mixing.
}
\label{fig:fermisurface}
\end{figure*}
We start with the low-energy theory for 
the clean AFQCM in two space dimensions,
\begin{widetext}
\begin{equation}
\begin{aligned}
& S_0
 = 
\sum_{N}
\int\dd {\bf k}~\psi^{\dagger}_{N,\sigma}({\bf k})\left\{ik_0
+
E_N[\vec k]
\right\}\psi_{N,\sigma}({\bf k})
+
\sum_{N}
\int \dd {\bf k}  \dd {\bf q} ~ 
g^{(N)}_{k_N+q_N, k_{\bar N}}
\psi^\dagger_{N,\sigma'}({\bf k}+{\bf q})
\vec \tau_{\sigma',\sigma}
\psi_{\overline{N},\sigma}({\bf k}) \cdot
\vec{\phi}({\bf q})
\\ &
+\frac{1}{4 \mu}\sum_{\{N_i\}}
\int \prod^{3}_{i=1} 
\dd {\bf k}_i~~
\lambda^{\{N_i \}; \{ \sigma_i \}}_{ \{k_{i;N_i}\}} ~
\psi^\dagger_{N_1,\sigma_1}({\bf k}_1)\psi^\dagger_{N_2,\sigma_2}({\bf k}_2)\psi_{N_3,\sigma_3}({\bf k}_3)\psi_{N_4,\sigma_4}(
{\bf k}_1+ {\bf k}_2- {\bf k}_3 
).
\end{aligned}
\label{eq:cleanaction}
\end{equation}
\end{widetext}
Here, $O(3)$ symmetric critical spin fluctuations are described by a three-component bosonic field $\vec \phi ({\bf q})$.
${\bf q} \equiv (q_0,\vec{q})$ is a frequency-momentum vector with $\dd{\bf q} = \frac{\dd q_0\dd q_x \dd q_y}{(2\pi)^3}$. 
$\vec q$ 
denotes the deviation from the antiferromagnetic wave-vector 
$\vec Q_{AF}$, 
which is chosen to be $2 \vec{Q}_{AF} = 0$ up to a reciprocal lattice vector. 
The bare kinetic term for the boson is dropped because its low-energy dynamics is entirely fixed by the interaction with electrons \cite{SCHLIEF}.
$\vec Q_{AF}$ connects four pairs of hot spots on the Fermi surface as is shown in \fig{fig:HS}.
Hence,
we divide the Fermi surface into eight disjoint segments, each containing one hot spot.
$\psi_{N,\sigma}({\bf k})$ represents the electron in segment $N$
with spin $\sigma$,
where
$\vec k$ represents the deviation from hot spot $N$.
All repeated spin indices are summed over.
In each segment, 
the momentum component $k_N$ 
perpendicular to $\vec Q_{AF}$ is 
used to label points on the Fermi surface within 
$-(k_F-\delta k_F) \leq k_N < k_F+\delta k_F$, 
where $2k_F$ is the
linear size of a segment.
Therefore, Fermi velocity and couplings are functions of $k_N$.
For $N=1$, $k_N=k_x$ and 
the electron dispersion 
is written as
$E_1[\vec k]
= V^{(1)}_{F,k_x} \left( v^{(1)}_{k_x} k_x + k_y \right)$,
where 
$V^{(1)}_{F,k_x}$ is the $k_x$-dependent Fermi velocity parallel to $\vec Q_{AF}$ 
and $v^{(1)}_{k_x}$ specifies the shape of Fermi surface through $E_1[\vec{k}] = 0$.
We choose the scale of frequency so that $V_{\mathrm{F},0} =1$.
$g^{(N)}_{k_N',k_{\bar N}}$
represents the coupling function that describes a low-energy electron scattered by the boson 
from $k_{\bar N}$ in segment $\overline{N}$ to $k'_{N}$ in segment $N$, 
where $\overline{N}$ represents the segment connected to $N$ through $\vec{Q}_{AF}$. 
In segment 1, we write those coupling functions without the superscript as
$g^{(1)}_{k',k}= g_{k',k}$, 
$V^{(1)}_{F,k}=V_{F,k}$
and
$v^{(1)}_{k} = v_{k}$.
The coupling functions in other segments are fixed by these functions through $D_4$ symmetry.
$v_{k}$ is referred to as nesting angle as pairs of segments connected by $\vec Q_{AF}$ become nested when $v_{k}=0$.
Finally, $\lambda^{\{N_i \}; \{ \sigma_i \}}_{ \{k_{i;N_i}\}}$ represents the four-fermion coupling function.
All coupling functions are defined by the vertex function evaluated at energy scale $\mu$\cite{BORGES2023169221}.
What is discussed in the present paper can be easily generalized to the $SU(N_c)$ symmetric theory with no essential change\cite{SCHLIEF,BORGES2023169221}, 
where the critical boson is generalized to 
$\Phi_{\sigma\sigma'} = \sum_{a=1}^{N_c^2-1} \phi^a 
\tau^a_{\sigma\sigma'}$
with
$\tau^a_{\sigma\sigma'}$ being the generators of $SU(N_c)$.
However, we will focus on the physically most relevant case with $N_c=2$.

We note that  ${\cal C} \equiv 
\left\{ V_{F,k}, v_{k},  g_{k',k}, \lambda^{\{N_i \}; \{ \sigma_i \}}_{ \{k_{i;N_i}\}} \right\}$ is the minimal and complete set of parameters that are needed to fully specify the low-energy physics.
Therefore, the infrared physics of the system is entirely encoded in the functional renormalization group (RG) flow of these coupling functions with lowering $\mu$.
In ${\cal C}$, 
an interacting fixed point arises at 
$V_{F,k}=1$,
$\lambda^{\{N_i \}; \{ \sigma_i \}}_{ \{k_{i;N_i}\}} =0$
and
$g_{k',k}^2/v_k = \pi/2$
in the $v_{k}\rightarrow 0$ limit \cite{BORGES2023169221}.
This is a non-perturbative fixed point at which the critical boson acquires an anomalous dimension $1$, and the dynamical critical exponent is $z=1$, which are exact\cite{SCHLIEF}. 
On the other hand, fermions do not have an anomalous dimension because the feedback of the strongly damped boson to the fermion is negligible at the fixed point. 
The behavior of theories tuned away from the fixed point is controlled by the bare nesting angle.
To the leading order in the nesting angle, the critical boson acquires an overdamped propagator,
$D({\bf q}) = \left[ \abs{q_0} + c(\abs{q_x}+\abs{q_y})\right]^{-1}$ with
speed 
$c = \sqrt{\frac{v_0}{16}\log\left(\frac{1}{v_0}\right)}$ set by the nesting angle at the hot spots.
Under the RG flow,
$v_0$ flows toward zero as
\begin{equation}
v_0(\ell) = \frac{\pi^2}{3}\frac{1}{(\ell+\ell_0)\log(\ell+\ell_0)},
\label{eq:c}
\end{equation}
where $\ell \equiv \log \Lambda/\mu$ is the logarithmic length scale associated with energy scale $\mu$
, $\Lambda$ is a UV cutoff,
and
$\ell_0 
\sim \frac{1}{v_0(0) \log 1/v_0(0)}
$ is the length scale set by the bare nesting angle.
For $v_0(0) \ll 1$, there is a large window of length scale $0 < \ell \ll \ell_0$ in which  $v_0(\ell) \approx v_0(0)$, and the flow of $v_0$ can be ignored.
Within this window of length scale, which becomes large for small $v_0(0)$, the physics is governed by a non-Fermi liquid quasi-fixed point with $v_0$-dependent critical exponents.
It is only a quasi-fixed point because $v_0$ is not exactly marginal, and the RG flow is cut off by a superconducting instability.
Superconductivity is driven by the four-fermion coupling generated from the critical spin fluctuations whose strength is of the order of  $g_{0,0}^4/c^2 \sim  v_0(0)/log (1/v_0(0))$ near the hot spots\cite{BORGES2023169221}. 
Eventually, the four-fermion coupling becomes large, and the superconducting instability arises at 
$T_c \sim \Lambda e^{-\frac{1}{\sqrt{v_0(0)}}}$ \cite{ BORGES2023169221, PhysRevB.95.174520}.
However, the four-fermion coupling is negligible at length scales $\ell \ll 
\frac{1}{\sqrt{v_0(0)}}$.
Here, we focus on the small $v_0$ limit and the window of energy scale above the superconducting transition temperature where the four-fermion coupling can be ignored.

Now, we add a magnetic impurity to this critical metal.
We represent the impurity spin as
$\vec{S}_{\mathrm{imp}} = 
\sum_{\alpha,\beta}
f^\dagger_\alpha
\frac{\vec{\tau}_{\alpha,\beta}}{2}f_{\beta}$, 
where $f_\alpha$ is  
a pseudo-fermion field with spin $\alpha$.
The dynamics of the  impurity is described by 
\begin{widetext}
\begin{equation}
\begin{aligned}
&  S_1  =
\int\frac{\dd p_0}{2\pi}\left(i p_0 +  i \frac{\pi}{2 \beta} \right) f^{\dagger}_{\alpha} (p_0)f_{\alpha}(p_0)
\\ &
 +
\int\dd{\bf q} ~
\left( 
\int  \frac{\dd p_0}{2\pi}
f^\dagger_\alpha(p_0)
\frac{\vec{\tau}_{\alpha,\beta}}{2}f_{\beta}(p_0+q_0)
\right)
\cdot \left( 
\sum_{N,N'} 
\int\dd{\bf k}
\frac{
J^{(N,N')}_{(k+q)_N,k_{N'}}
}{\mu} 
\psi^{\dagger}_{N,\sigma}({\bf k}+{\bf q})\frac{\vec{\tau}_{\sigma,\sigma'}}{2}\psi_{N',\sigma'}({\bf k})
 + g_f
\vec \phi(-{\bf q})
\right).
\end{aligned}
\label{eq:ImpurityAction}
\end{equation}
\end{widetext}
The imaginary chemical potential projects out states with $f^\dagger_\alpha f_\alpha = 0,2$, 
where $\beta=1/(k_BT)$ becomes infinity at zero temperature \cite{popov1988functional}.  
$J^{(N,N')}_{k_N,k'_{N'}}$ is the Kondo coupling function that describes the scattering of electrons from $(N',k_{N'}')$ to $(N,k_N)$ through interaction with the impurity.
$g_f$ represents the boson-impurity coupling (see Fig. \ref{fig:setup}).
At the impurity-free fixed point,
the fields have scaling dimensions $[\Psi({\bf k})]=[\vec \phi({\bf k})] =-2$ with respect to the transformation that rescales both components of momentum with dimension
$[{\bf  k}]=1$.
Under this scale transformation,
the dimensionless size of Fermi surface ($k_F$), measured in the unit of the floating energy scale ($\mu$) that is sent to zero in the low-energy limit, grows, playing the role of a relevant parameter \cite{
BORGES2023169221,
PhysRevB.110.155142}. 
$g_f$ and $J$
have dimensions $0$ and $-1$, respectively.
Accordingly, the power of $\mu$ has been factored out in each coupling.
It is noted that $g_f$ is marginal due to the large anomalous dimension of the boson generated at the strongly coupled clean fixed point.
Without it, $g_f$ would be strictly relevant \cite{Varma_Kondo_effect_AFM_2005}.

While the Kondo coupling has the negative scaling dimension,
we have to keep it within the low-energy theory, as it gives rise to a logarithmic divergence.
Due to scatterings that involve large momentum transfer, quantum corrections are proportional to $k_F$, 
which alters the relevancy of a coupling from what is expected from its scaling dimension.
%
Kondo coupling is particularly prone to this 
large-momentum/low-energy (in short, UV/IR) mixing \cite{IPSITA,
PhysRevLett.128.106402, BORGES2023169221, PhysRevB.110.155142}
because the phase space available for low-energy electrons becomes even greater without the momentum conservation\footnote{Another such coupling is the four-fermion coupling}.
To understand this, let us first consider the beta functional of the Kondo coupling in 
the absence of critical boson:
$\frac{\dd 
{J}^{(N,N')}_{k, k'}}{\dd \ell} = 
-{J}^{(N,N')}_{k, k'}
+
\sum_{M}
\frac{1}{2\pi}
\int_{-k_F}^{k_F}
\frac{\dd q}{2\pi \mu V_{F,q}}
{J}^{(N,M)}_{k,q}{J}^{(M,N')}_{q,k'}.
$
Unlike the usual beta function used in the literature, $-J^{(N,N')}_{k,k'}$ term appears because the scaling dimension of the Kondo coupling is $-1$ under the scaling in which both components of momentum parallel and perpendicular to the Fermi surface are scaled in the same way.
We choose this scheme to keep the dynamics of the critical boson invariant under the scale transformation. 
In the scheme used for Fermi liquids in the literature, only the momentum perpendicular to the Fermi surface is rescaled, and the Kondo coupling is deemed marginal.
However, the physics does not depend on the scaling scheme.
The difference in the tree-level scaling dimension is compensated by the fact that $k_F$ that sets the cutoff of momentum along the Fermi surface has scaling dimension $1$ in the present scheme, and grows under the scale transformation.
The interaction that scatters an electron from $k'$ to $k$ is renormalized by a virtual electron that can be placed
anywhere on the Fermi surface (hence, $q$ and $M$ are integrated and summed over).
As the volume of the phase space is controlled by the relevant parameter $k_F$,
the expansion is actually controlled by a dimensionless parameter 
$\tilde J \sim \int \dd q J \sim k_F J$, 
which is marginally relevant for $J>0$ \cite{10.1143/PTP.32.37}.

In the AFQCM, the beta function is modified by critical spin fluctuations,
and is augmented with the beta function for the boson-impurity coupling.
To the leading order in $v$, $J$ and $g_f$,
the beta functions become
(see Appendices \ref{sec:vertexfunction} 
and
\ref{sec:quantumcorrections}
for details)
\begin{align}
&\begin{aligned}
\frac{\dd {J}^{(N,N')}_{k, k'}}{\dd \ell} &= 
- 
\left(2-z
+
\eta_f
\right)
{J}^{(N,N')}_{k,k'} 
\\ & + 
\sum_{M}
\frac{1}{2\pi} 
\int\frac{\dd q}{2\pi \mu V_{F,q}}
{J}^{(N,M)}_{k,q}
{J}^{(M,N')}_{q,k'}, 
\end{aligned}\label{eq:betaJ}
\\
&\frac{\partial g_f }{\partial \ell} = -  \left(
    \eta_f 
    + \eta^{(\Phi)}\right) g_f,
    \label{eq:betag}
\end{align}
where
$z=1+\frac{3}{4\pi} 
\frac{v_0}{c}$ 
is the dynamical critical exponent corrected from $1$ by the non-zero nesting angle
and
$\eta_f = \frac{\tilde g_f}{\pi^3} \ell$
with
$\tilde g_f \equiv
\frac{g_f^2}{c^2}$.
Both $g_f$ and $J$ 
have the common component of 
anomalous dimension 
$\eta_f$,
which arises from 
the self-energy and vertex correction of pseudo-fermion.
The flow of $g_f$ is also affected by the finite $v$-correction to the anomalous dimension of the boson
,
$\eta^{(\Phi)} =  \frac{1}{4\pi} 
\frac{v_0}{c}
\log
\frac
{c}
{v_0}$ \cite{SCHLIEF}.
Acting on top of the anomalous dimension $1$ of the boson at the $v=0$ fixed point,
$\etaphi$ makes the boson-impurity coupling slightly irrelevant at a small but non-zero nesting angle.
Without the anomalous dimension of the boson $1$ that arises at the interacting fixed point, the beta function for $g_f$ would be
$\frac{\partial g_f }{\partial \ell} \sim   g_f$
and $g_f$ would become large at low energies.
With higher-order corrections included in the beta function of $g_f$,
a fixed point with $g_f \sim O(1)$ and $J = 0$ may become stable.
In the AFQCM, however, this does not happen because the large anomalous dimension of the boson renders the boson-impurity coupling a marginally irrelevant coupling \cite{PhysRevB.66.024426,PhysRevB.66.024427,PhysRevB.61.4041}.
The self-energy and vertex correction for itinerant electrons would generate additional contributions to the beta functional of $J$.
However, those corrections have been dropped.
They are not important at low energies 
because critical spin fluctuations renormalize itinerant electrons only within patches of size $\mu/(v_0 c)$ around the hot spots at energy scale $\mu$ \cite{BORGES2023169221}.
This can be understood as follows.
An electron on the Fermi surface at momentum $k_N$ away from hot spot $N$ must interact with a boson with minimum momentum $q \sim v_0k_N$ 
(and energy $\sim cv_0k_N$) to be scattered onto the Fermi surface in segment $\bar N$
due to an imperfect nesting for $v_0 \neq 0$.
So, electrons with $k_N \gg \mu/(v_0c)$ decouple from spin fluctuations at energies lower than $\mu$.
Since the size of the `hot patches' shrinks linearly in $\mu$, those quantum corrections do not affect ${J}^{(N,N')}_{k, k'}$ for most $k, k'$.
Therefore, quantum corrections to those hot electrons are suppressed by $\mu/k_F$ for the s-wave Kondo coupling.
On the contrary, the 
 pseudo-fermion self-energy and the vertex correction for the impurity-boson coupling
 gives rise to the anomalous dimension $\eta_f$
 for  ${J}^{(N,N')}_{k, k'}$ regardless of $k$ and $k'$.
Remarkably, 
$\eta_f = \tgf \ell/\pi^3$ 
 depends on 
$\ell = \log \Lambda/\mu$ explicitly.
This unusual sensitivity on the UV cutoff is traced back to the propagator of the boson at the impurity site, 
$\bar{D}(q_0) = \int \frac{\dd q_x \dd q_y}{(2\pi)^2}D({\bf q}) \sim 
-\abs{q_0}
\log\left(\frac{\Lambda}{\abs{q_0}}\right)$.
The UV divergence in the term non-analytic in $q_0$ implies that short-wavelength bosons significantly contribute to the long-time correlation.
Namely, short-wavelength spin textures whose wave-vectors deviate significantly from $\vec Q_{AF}$ still fluctuate slowly temporally.
This is caused by the strong mixing of the boson with low-energy particle-hole excitations that carry large momenta.
The low-energy spectral weight enhanced by large-momentum modes also gives rise to a logarithmically divergent specific heat coefficient from the boson\cite{SCHLIEF}.
This {\it UV/IR mixing} plays the crucial role in determining the fate of the Kondo coupling at low energies.
Eqs. \eqref{eq:betaJ} and  \eqref{eq:betag} are valid for $\ell > \ell_i \sim O(1)$ 
where this UV/IR-driven contribution
is dominant over terms that are $O(\ell^0)$ in $\eta_f$.

To simplify the task of solving the beta functional,
we focus on the 
`s-wave' channel of the form
${J}^{(N,N')}_{k, k'} =  
\frac{\pi^2}{4} 
\frac{\mu}{k_F} 
\sqrt{V^{(N)}_{F,k}V^{(N')}_{F,k'}} \tilde J^V$,
where ${J}^{(N,N')}_{k, k'}$ is momentum-independent modulo the factor of Fermi velocity.
Fermi velocity outside the hot patches increases as 
$\frac{\partial V_{F,k}}{\partial \ell} =  (z-1) V_{F,k}$
relative to $V_{F,0}$, which is set to be $1$, 
because only electrons within the hot patches are renormalized by spin fluctuations \cite{BORGES2023169221}.
The s-wave Kondo coupling $\tilde J^V$,
which includes 
the effect of growing phase space
$k_F/\mu$,
satisfies a simple beta function,
\begin{equation}
\frac{\partial \tilde J^V(\ell)}{\partial \ell} =  
-
\eta_{f}(\ell) 
\tilde{J}^V(\ell) +
\left(\tilde{J}^V(\ell)\right)^2,
\label{eq5}
\end{equation}
where
$\eta_{f}
=
\frac{\tgf}{\pi^3} \ell$.
In this problem, there are three parameters that are marginal up to logarithmic corrections:
the nesting angle at the hot spots ($v_0$),
Kondo coupling ($\tJV$),
and the boson-impurity coupling ($g_f$).
Let
$v_{0,i} \sim 
1/(\ell_0 \log \ell_0)$,
$\tJV_i$ and $g_{f,i}$ 
denote 
the nesting angle,
Kondo coupling
and boson-impurity coupling, 
respectively,
defined at short-distance cutoff scale $\ell_i \sim O(1)$.
Our main goal is to understand how the critical spin fluctuations affect the behavior of Kondo coupling.
In particular, we would like to extract how the Kondo scale, the scale at which $\tJV(\ell)$ becomes large, depends on 
$v_{0,i}$ and $\tgfi$
in the small $\tJV_i$ limit.
Our calculation is controlled in the limit that $v_{0,i}$, $\tJV$ and $\tgf$ are small without a particular order among them.

We first solve \eq{eq:betag}.  
Since $\etaphi_i$ and $\tgfi \ell$ set the rates at which $g_f$ decays with increasing $\ell$,
we define the following characteristic scales:
$1/\etaphi_i \sim \sez$ 
and
$\ell_1 \sim 
1/\sqrt{\tgfi}$.
Naturally, the RG flow exhibits different behaviors, depending on the relative magnitude of $1/\ell_0$ and $\tgfi$.

If $\tgfi \ll 1/\ell_0$,
the effect of the boson-impurity coupling is weak compared to the quantum correction that arises from a non-zero nesting angle.
Since $\etaphi$ plays the dominant role over $\tgf$ in determining the flow of $\tgf$,
one can ignore $\tgf$ in \eq{eq:betag}
to obtain
$g_f(\ell) =  e^{-\frac{\sqrt{\ell +\ell_0}-\sqrt{\ell_0+\ell_i}}{\sqrt{3}}}g_{f,i}$.
In this case, the boson-impurity coupling decays to zero exponentially at large $\ell$
due to the relatively large correction to the anomalous dimension of the boson,
and
$\tgf$ remains negligible for the flow of $g_f$ at all scales.
With small $\tgf$,
one essentially recovers the Kondo effect in Fermi liquids: 
$\ell_K \sim 1/\tJV_i$ with a small correction.

\begin{figure}[t]
    \centering
    \includegraphics[width=\linewidth]{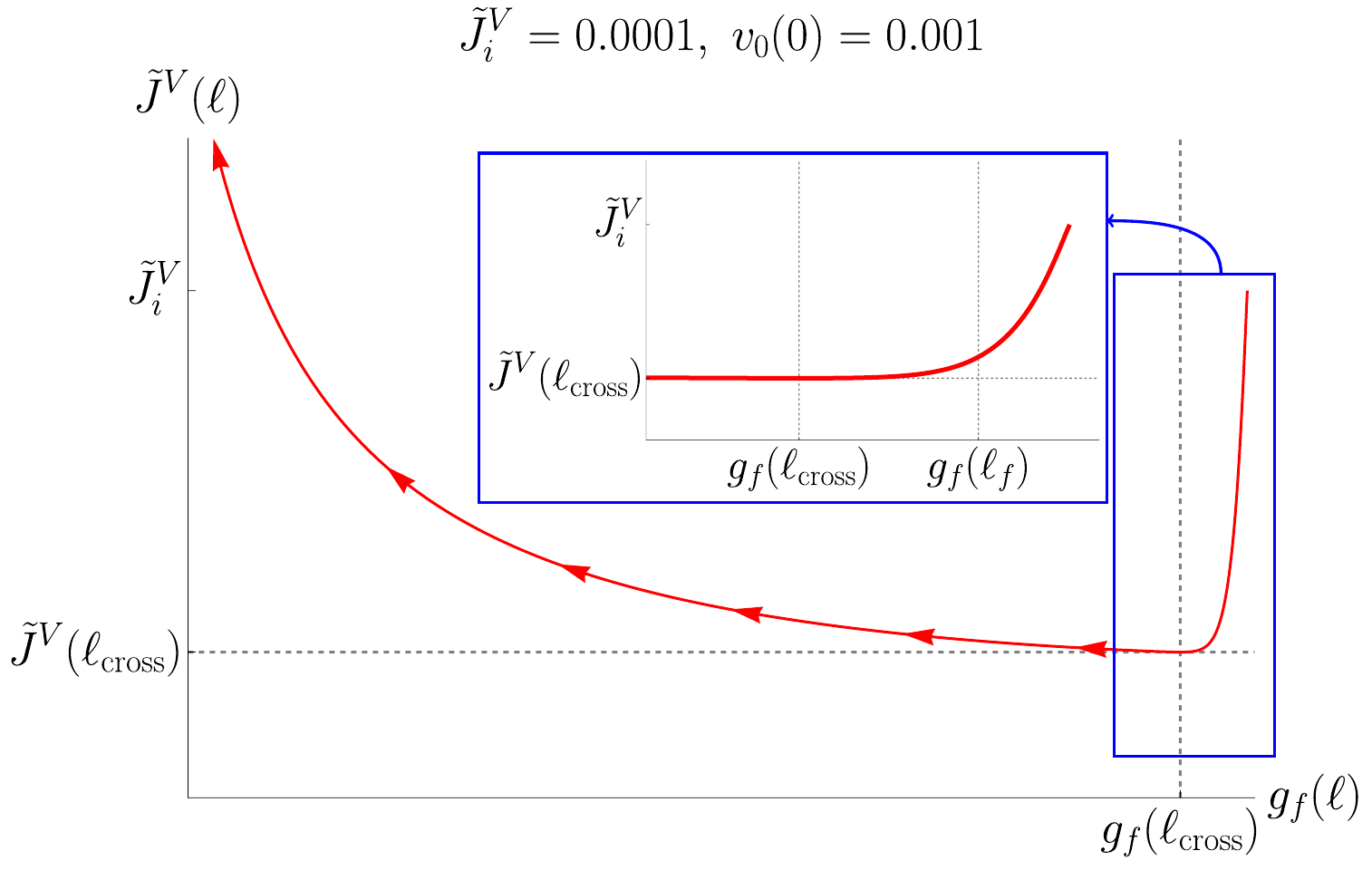}
    \caption{
    RG flow of the $\tilde{J}^V$ and $g_f$. 
    The blue box zooms into the RG flow at short distances.
    For $\ell \leq \ell_f$, the large boson-impurity coupling suppresses the Kondo coupling, making it decrease with increasing length scale.
    For $\ell > \ell_f$, the boson-impurity coupling becomes subdominant, and the Kondo coupling begins to grow as in Fermi liquids.
    However, the suppressed Kondo coupling at $\ell_f$ greatly increases the length scale below which the Kondo coupling becomes $O(1)$.
    }
    \label{fig:kondoflowfig}
\end{figure}

A qualitatively new behavior emerges for
$\tgfi \gg 1/\ell_0$. 
In this case, the bare boson-impurity coupling is strong enough that 
$\etaphi$ is negligible
in \eq{eq:betag}
at short distance scales.
The $\tgf$-dominated flow gives 
$\tgf(\ell) = \frac{3 \pi^3 \tilde{g}_{f,i}(\ell+\ell_0)}{3 \pi^3 (\ell_0+\ell_i)+\tilde{g}_{f,i}  \left[2\ell^3+3\ell^2\ell_0-\ell_i^2(3\ell_0+2\ell_i)\right]}$. 
This slowly decaying $\tgf$ gives rise to 
a relatively large anomalous dimension,
which suppresses the Kondo coupling significantly.
To understand this quantitatively
(see Appendix \ref{sec:qptkondo} for details), 
we note that
$\tgf \ell/\pi^3 \approx 1/\ell$
within length scale $\ell_1 \ll \ell \ll \ell_0$,
and \eq{eq5} becomes
$\frac{\partial \tilde J^V}{\partial \ell} =  
-\tilde{J}^V/\ell +
(\tilde{J}^V)^2$.
Its solution is given by
$\tilde{J}^V(\ell)^{-1} \approx 
\ell \left[
\frac{1}{\tJV(\ell_1) \ell_1}
- \log \frac{\ell}{\ell_1}
\right]
$.
According to this,
Kondo coupling only grows as a logarithm of $\ell$,
and the Kondo scale becomes
$\ell_K = \ell_1 e^{
\frac{1}{\tJV(\ell_1) \ell_1}
}$.
This implies that the Kondo temperature $T_K = \Lambda e^{-\ell_K}$ is suppressed by an exponential of an exponential of $1/\tJV(\ell_1)$.
In a sense, the critical boson demotes Kondo coupling from a marginally relevant coupling to a `marginally marginally relevant' coupling.
However, this rapid increase of $\ell_K$ 
with decreasing $\tJV_i$ 
is cut off once 
$\ell_K$ becomes large enough that one can not drop $\etaphi$ 
in \eq{eq:betag} anymore.
Because 
$\eta_f = \tgf \ell/\pi^3 \sim 1/\ell$
decays faster than $\etaphi \sim 1/(\ell+\ell_0)^{1/2}$,
$g_f$ exhibits 
a crossover from the $\tgf$-dominated flow to the $\etaphi$-dominated flow around
$\ell_f \sim \sqrt{\ell_0}$.
For $\ell > \ell_f$, the flow of $g_f$ is dominated by $\etaphi$ and decays exponentially.
As the anomalous dimension for $\tJV$ becomes exponentially small for $\ell > \ell_f$,
the flow of Kondo coupling is reduced to that of the Fermi liquids.
The Kondo scale at which $J$ becomes $O(1)$ is given by
$\ell_K \sim \ell_f + 
\frac{1}{\tJV(\ell_f)}$,
where 
$\tJV(\ell_f) \sim 
\tJV_i
\frac{
v_{0,i} \log 1/v_{0,i}}{
g_{f,i}}
$
is the Kondo coupling at scale $\ell_f$.  
In the small $\tJV_i$ limit,
$
\frac{1}{\tJV(\ell_f)} \gg \ell_f$
and
the Kondo scale becomes
$\ell_K \sim 
\frac{1}{\tJV_i}
\frac{ g_{f,i}  }{v_{0,i} \log 1/v_{0,i}}
$ to the leading order in $\tJV_i$.
The interplay between the impurity-boson coupling and the Kondo coupling is illustrated in Fig. \ref{fig:kondoflowfig}.

\begin{figure}[t]
\centering
\includegraphics[width=\linewidth]{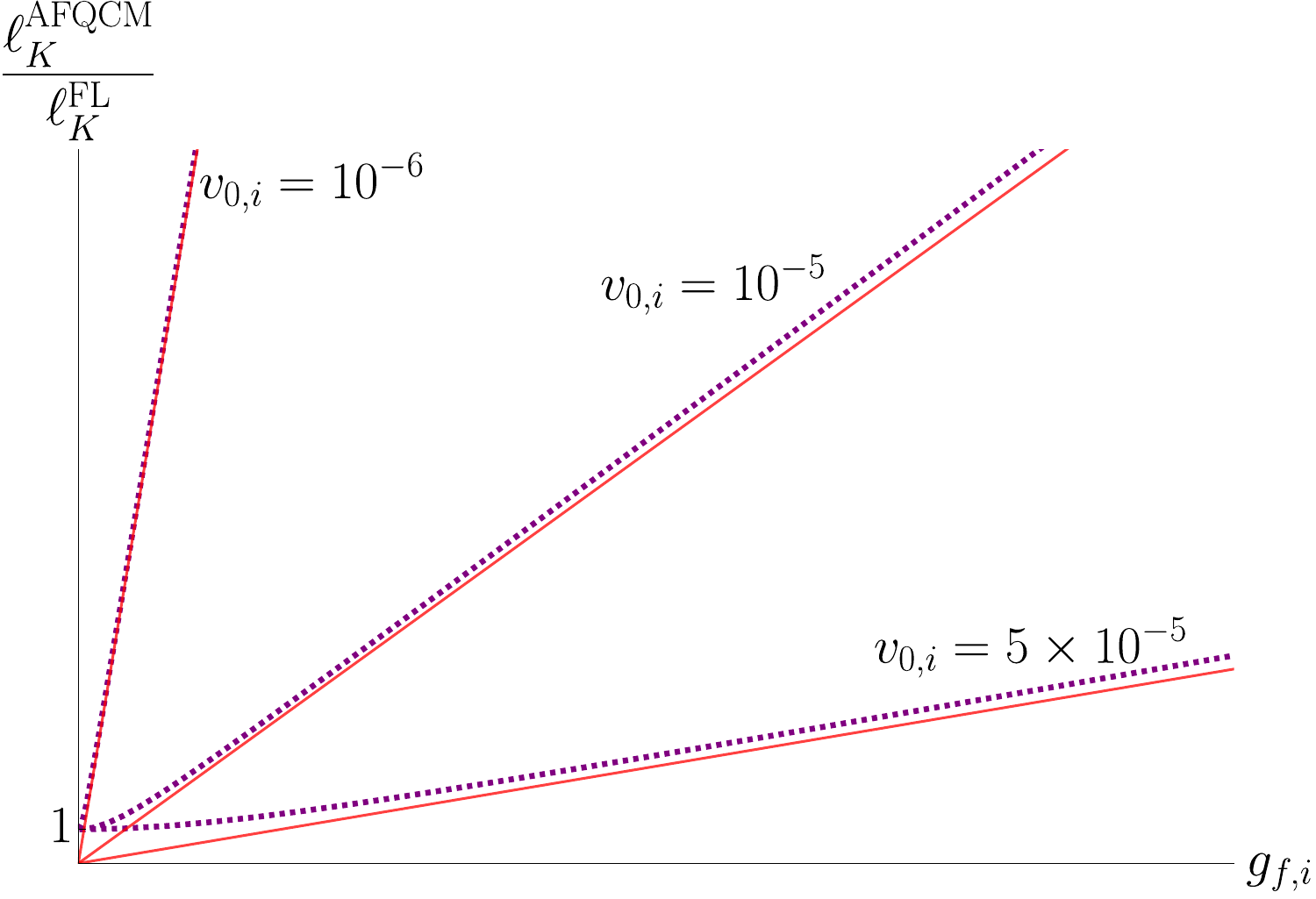}
    \caption{
Logarithmic Kondo scale of the AFQCM relative to that of the Fermi liquid with the same electronic density of state and bare Kondo coupling plotted 
as a function of the bare boson-impurity coupling, $g_{f,i}$ 
for three different bare nesting angles $v_{0,i}$.
The dashed lines represent the numerical solutions of Eqs. \eqref{eq:betag} 
and \eqref{eq5} 
obtained with $\tJV_i = 10^{-8}$,
and the solid lines 
are
$\frac{\ell^{\mathrm{AFQCM}}}{\ell^{\mathrm{FL}}}=
A \frac{g_{f,i}}{v_{0,i} \log (1/v_{0,i})}$
with
$A = \frac{8 e^{3/4}}{3\sqrt{\pi }}$.
}
\label{fig:phasediagram}
\end{figure}

At small $v_{0,i}$,
$\ell_K$ is enhanced by the factor of
$\frac{ g_{f,i}  }{v_{0,i} \log 1/v_{0,i}}$
compared to that of Fermi liquid ($\ell_K^{FL} \sim 1/\tJV_i$) with the same density of state (see Fig. \ref{fig:singlet}).
This is confirmed through solving the beta functions numerically as is shown in \fig{fig:phasediagram}.
For $T > \Lambda e^{-\ell_f}$, the Kondo coupling decreases with decreasing temperature, and the resistivity decreases with decreasing temperature.
One expects to see a logarithmic upturn of resistivity only in temperatures below $\Lambda e^{-\ell_f}$ in which the Kondo coupling begins to grow with decreasing temperature.
However, the Kondo screening occurs at a much lower temperature due to the suppression of the renormalized Kondo coupling at scale $\ell_f$.
In the small $v_{0,i}$ limit, the Kondo length scale becomes larger than the superconducting length scale given by
$\ell_{SC} \sim 1/\sqrt{v_{0,i} \log 1/v_{0,i}}$ \cite{BORGES2023169221}, and the logarithmic upturn of the resistivity becomes unobservable in the normal state.

%

In summary, the Kondo screening is greatly suppressed in AFQCM through a UV/IR mixing
when the hot spots on the Fermi surface are well nested. 
Our analysis fully takes into account  the strong electron-boson coupling in the limit that the nesting angle is small, 
but it is still perturbative in the Kondo coupling.
In the future, it is of great interest to 
treat the Kondo coupling non-perturbatively as well \cite{Wilson_RMP,Andrei_1980,Andrei_Destri_1984,AFFLECK1990517,AFFLECK1991849}.

\section*{Acknowledgement}

This research was supported by the Natural Sciences 
and Engineering Research Council of
Canada. Research at the Perimeter Institute is supported in part by the
Government of Canada through Industry Canada, and the Province of
Ontario through the Ministry of Research and Information. P.L. is supported by the U.S. National Science Foundation grant No. DMR-2245246, by the Harvard Quantum Initiative Postdoctoral Fellowship in Science and Engineering, and by the Simons Collaboration on Ultra-Quantum Matter which is a grant from the Simons Foundation (651440, Subir Sachdev).

\bibliography{references}

\begin{thebibliography}{43}%
\makeatletter
\providecommand \@ifxundefined [1]{%
 \@ifx{#1\undefined}
}%
\providecommand \@ifnum [1]{%
 \ifnum #1\expandafter \@firstoftwo
 \else \expandafter \@secondoftwo
 \fi
}%
\providecommand \@ifx [1]{%
 \ifx #1\expandafter \@firstoftwo
 \else \expandafter \@secondoftwo
 \fi
}%
\providecommand \natexlab [1]{#1}%
\providecommand \enquote  [1]{``#1''}%
\providecommand \bibnamefont  [1]{#1}%
\providecommand \bibfnamefont [1]{#1}%
\providecommand \citenamefont [1]{#1}%
\providecommand \href@noop [0]{\@secondoftwo}%
\providecommand \href [0]{\begingroup \@sanitize@url \@href}%
\providecommand \@href[1]{\@@startlink{#1}\@@href}%
\providecommand \@@href[1]{\endgroup#1\@@endlink}%
\providecommand \@sanitize@url [0]{\catcode `\\12\catcode `\$12\catcode
  `\&12\catcode `\#12\catcode `\^12\catcode `\_12\catcode `\%12\relax}%
\providecommand \@@startlink[1]{}%
\providecommand \@@endlink[0]{}%
\providecommand \url  [0]{\begingroup\@sanitize@url \@url }%
\providecommand \@url [1]{\endgroup\@href {#1}{\urlprefix }}%
\providecommand \urlprefix  [0]{URL }%
\providecommand \Eprint [0]{\href }%
\providecommand \doibase [0]{https://doi.org/}%
\providecommand \selectlanguage [0]{\@gobble}%
\providecommand \bibinfo  [0]{\@secondoftwo}%
\providecommand \bibfield  [0]{\@secondoftwo}%
\providecommand \translation [1]{[#1]}%
\providecommand \BibitemOpen [0]{}%
\providecommand \bibitemStop [0]{}%
\providecommand \bibitemNoStop [0]{.\EOS\space}%
\providecommand \EOS [0]{\spacefactor3000\relax}%
\providecommand \BibitemShut  [1]{\csname bibitem#1\endcsname}%
\let\auto@bib@innerbib\@empty
\bibitem [{\citenamefont {Anderson}(1961)}]{PhysRev.124.41}%
  \BibitemOpen
  \bibfield  {author} {\bibinfo {author} {\bibfnamefont {P.~W.}\ \bibnamefont
  {Anderson}},\ }\bibfield  {title} {\bibinfo {title} {Localized magnetic
  states in metals},\ }\href {https://doi.org/10.1103/PhysRev.124.41}
  {\bibfield  {journal} {\bibinfo  {journal} {Phys. Rev.}\ }\textbf {\bibinfo
  {volume} {124}},\ \bibinfo {pages} {41} (\bibinfo {year} {1961})}\BibitemShut
  {NoStop}%
\bibitem [{\citenamefont {Kondo}(1964)}]{10.1143/PTP.32.37}%
  \BibitemOpen
  \bibfield  {author} {\bibinfo {author} {\bibfnamefont {J.}~\bibnamefont
  {Kondo}},\ }\bibfield  {title} {\bibinfo {title} {Resistance minimum in
  dilute magnetic alloys},\ }\href {https://doi.org/10.1143/PTP.32.37}
  {\bibfield  {journal} {\bibinfo  {journal} {Progress of Theoretical Physics}\
  }\textbf {\bibinfo {volume} {32}},\ \bibinfo {pages} {37} (\bibinfo {year}
  {1964})},\ \Eprint
  {https://arxiv.org/abs/https://academic.oup.com/ptp/article-pdf/32/1/37/5193092/32-1-37.pdf}
  {https://academic.oup.com/ptp/article-pdf/32/1/37/5193092/32-1-37.pdf}
  \BibitemShut {NoStop}%
\bibitem [{\citenamefont {Nozi{\`e}res}(1974)}]{Nozieres:1974aa}%
  \BibitemOpen
  \bibfield  {author} {\bibinfo {author} {\bibfnamefont {P.}~\bibnamefont
  {Nozi{\`e}res}},\ }\bibfield  {title} {\bibinfo {title} {A
  ``fermi-liquid''description of the kondo problem at low temperatures},\
  }\href {https://doi.org/10.1007/BF00654541} {\bibfield  {journal} {\bibinfo
  {journal} {Journal of Low Temperature Physics}\ }\textbf {\bibinfo {volume}
  {17}},\ \bibinfo {pages} {31} (\bibinfo {year} {1974})}\BibitemShut {NoStop}%
\bibitem [{\citenamefont {Doniach}(1977)}]{DONIACH1977231}%
  \BibitemOpen
  \bibfield  {author} {\bibinfo {author} {\bibfnamefont {S.}~\bibnamefont
  {Doniach}},\ }\bibfield  {title} {\bibinfo {title} {The kondo lattice and
  weak antiferromagnetism},\ }\href
  {https://doi.org/https://doi.org/10.1016/0378-4363(77)90190-5} {\bibfield
  {journal} {\bibinfo  {journal} {Physica B+C}\ }\textbf {\bibinfo {volume}
  {91}},\ \bibinfo {pages} {231} (\bibinfo {year} {1977})}\BibitemShut
  {NoStop}%
\bibitem [{\citenamefont {Hewson}(1993)}]{Hewson1993}%
  \BibitemOpen
  \bibfield  {author} {\bibinfo {author} {\bibfnamefont {A.~C.}\ \bibnamefont
  {Hewson}},\ }\href {https://doi.org/doi.org/10.1017/CBO9780511470752} {\emph
  {\bibinfo {title} {The Kondo Problem to Heavy Fermions}}},\ \bibinfo {series}
  {Cambridge Studies in Magnetism}, Vol.~\bibinfo {volume} {2}\ (\bibinfo
  {publisher} {Cambridge University Press},\ \bibinfo {address} {Cambridge},\
  \bibinfo {year} {1993})\BibitemShut {NoStop}%
\bibitem [{\citenamefont {Stewart}(1984)}]{RevModPhys.56.755}%
  \BibitemOpen
  \bibfield  {author} {\bibinfo {author} {\bibfnamefont {G.~R.}\ \bibnamefont
  {Stewart}},\ }\bibfield  {title} {\bibinfo {title} {Heavy-fermion systems},\
  }\href {https://doi.org/10.1103/RevModPhys.56.755} {\bibfield  {journal}
  {\bibinfo  {journal} {Rev. Mod. Phys.}\ }\textbf {\bibinfo {volume} {56}},\
  \bibinfo {pages} {755} (\bibinfo {year} {1984})}\BibitemShut {NoStop}%
\bibitem [{\citenamefont {Leggett}\ \emph {et~al.}(1987)\citenamefont
  {Leggett}, \citenamefont {Chakravarty}, \citenamefont {Dorsey}, \citenamefont
  {Fisher}, \citenamefont {Garg},\ and\ \citenamefont
  {Zwerger}}]{RevModPhys.59.1}%
  \BibitemOpen
  \bibfield  {author} {\bibinfo {author} {\bibfnamefont {A.~J.}\ \bibnamefont
  {Leggett}}, \bibinfo {author} {\bibfnamefont {S.}~\bibnamefont
  {Chakravarty}}, \bibinfo {author} {\bibfnamefont {A.~T.}\ \bibnamefont
  {Dorsey}}, \bibinfo {author} {\bibfnamefont {M.~P.~A.}\ \bibnamefont
  {Fisher}}, \bibinfo {author} {\bibfnamefont {A.}~\bibnamefont {Garg}},\ and\
  \bibinfo {author} {\bibfnamefont {W.}~\bibnamefont {Zwerger}},\ }\bibfield
  {title} {\bibinfo {title} {Dynamics of the dissipative two-state system},\
  }\href {https://doi.org/10.1103/RevModPhys.59.1} {\bibfield  {journal}
  {\bibinfo  {journal} {Rev. Mod. Phys.}\ }\textbf {\bibinfo {volume} {59}},\
  \bibinfo {pages} {1} (\bibinfo {year} {1987})}\BibitemShut {NoStop}%
\bibitem [{\citenamefont {Si}\ and\ \citenamefont
  {Steglich}(2010)}]{Si_Steglich_heavy_fermions_review}%
  \BibitemOpen
  \bibfield  {author} {\bibinfo {author} {\bibfnamefont {Q.}~\bibnamefont
  {Si}}\ and\ \bibinfo {author} {\bibfnamefont {F.}~\bibnamefont {Steglich}},\
  }\bibfield  {title} {\bibinfo {title} {Heavy fermions and quantum phase
  transitions},\ }\href {https://doi.org/10.1126/science.1191195} {\bibfield
  {journal} {\bibinfo  {journal} {Science}\ }\textbf {\bibinfo {volume}
  {329}},\ \bibinfo {pages} {1161} (\bibinfo {year} {2010})},\ \Eprint
  {https://arxiv.org/abs/https://www.science.org/doi/pdf/10.1126/science.1191195}
  {https://www.science.org/doi/pdf/10.1126/science.1191195} \BibitemShut
  {NoStop}%
\bibitem [{\citenamefont {Parcollet}\ \emph {et~al.}(1998)\citenamefont
  {Parcollet}, \citenamefont {Georges}, \citenamefont {Kotliar},\ and\
  \citenamefont {Sengupta}}]{PhysRevB.58.3794}%
  \BibitemOpen
  \bibfield  {author} {\bibinfo {author} {\bibfnamefont {O.}~\bibnamefont
  {Parcollet}}, \bibinfo {author} {\bibfnamefont {A.}~\bibnamefont {Georges}},
  \bibinfo {author} {\bibfnamefont {G.}~\bibnamefont {Kotliar}},\ and\ \bibinfo
  {author} {\bibfnamefont {A.}~\bibnamefont {Sengupta}},\ }\bibfield  {title}
  {\bibinfo {title} {Overscreened multichannel $\mathrm{SU}(n)$ kondo model:
  Large-$n$ solution and conformal field theory},\ }\href
  {https://doi.org/10.1103/PhysRevB.58.3794} {\bibfield  {journal} {\bibinfo
  {journal} {Phys. Rev. B}\ }\textbf {\bibinfo {volume} {58}},\ \bibinfo
  {pages} {3794} (\bibinfo {year} {1998})}\BibitemShut {NoStop}%
\bibitem [{\citenamefont {Senthil}\ \emph {et~al.}(2003)\citenamefont
  {Senthil}, \citenamefont {Sachdev},\ and\ \citenamefont
  {Vojta}}]{Senthil_Sachdev_Vojta_2003}%
  \BibitemOpen
  \bibfield  {author} {\bibinfo {author} {\bibfnamefont {T.}~\bibnamefont
  {Senthil}}, \bibinfo {author} {\bibfnamefont {S.}~\bibnamefont {Sachdev}},\
  and\ \bibinfo {author} {\bibfnamefont {M.}~\bibnamefont {Vojta}},\ }\bibfield
   {title} {\bibinfo {title} {Fractionalized fermi liquids},\ }\href
  {https://doi.org/10.1103/PhysRevLett.90.216403} {\bibfield  {journal}
  {\bibinfo  {journal} {Phys. Rev. Lett.}\ }\textbf {\bibinfo {volume} {90}},\
  \bibinfo {pages} {216403} (\bibinfo {year} {2003})}\BibitemShut {NoStop}%
\bibitem [{\citenamefont {Senthil}\ \emph {et~al.}(2004)\citenamefont
  {Senthil}, \citenamefont {Vojta},\ and\ \citenamefont
  {Sachdev}}]{Senthil_Vojta_Sachdev_2004}%
  \BibitemOpen
  \bibfield  {author} {\bibinfo {author} {\bibfnamefont {T.}~\bibnamefont
  {Senthil}}, \bibinfo {author} {\bibfnamefont {M.}~\bibnamefont {Vojta}},\
  and\ \bibinfo {author} {\bibfnamefont {S.}~\bibnamefont {Sachdev}},\
  }\bibfield  {title} {\bibinfo {title} {Weak magnetism and non-fermi liquids
  near heavy-fermion critical points},\ }\href
  {https://doi.org/10.1103/PhysRevB.69.035111} {\bibfield  {journal} {\bibinfo
  {journal} {Phys. Rev. B}\ }\textbf {\bibinfo {volume} {69}},\ \bibinfo
  {pages} {035111} (\bibinfo {year} {2004})}\BibitemShut {NoStop}%
\bibitem [{\citenamefont {Dzero}\ \emph {et~al.}(2010)\citenamefont {Dzero},
  \citenamefont {Sun}, \citenamefont {Galitski},\ and\ \citenamefont
  {Coleman}}]{Coleman_topological_kondo_ins}%
  \BibitemOpen
  \bibfield  {author} {\bibinfo {author} {\bibfnamefont {M.}~\bibnamefont
  {Dzero}}, \bibinfo {author} {\bibfnamefont {K.}~\bibnamefont {Sun}}, \bibinfo
  {author} {\bibfnamefont {V.}~\bibnamefont {Galitski}},\ and\ \bibinfo
  {author} {\bibfnamefont {P.}~\bibnamefont {Coleman}},\ }\bibfield  {title}
  {\bibinfo {title} {Topological kondo insulators},\ }\href
  {https://doi.org/10.1103/PhysRevLett.104.106408} {\bibfield  {journal}
  {\bibinfo  {journal} {Phys. Rev. Lett.}\ }\textbf {\bibinfo {volume} {104}},\
  \bibinfo {pages} {106408} (\bibinfo {year} {2010})}\BibitemShut {NoStop}%
\bibitem [{\citenamefont {Si}\ \emph {et~al.}(2014)\citenamefont {Si},
  \citenamefont {Pixley}, \citenamefont {Nica}, \citenamefont {Yamamoto},
  \citenamefont {Goswami}, \citenamefont {Yu},\ and\ \citenamefont
  {Kirchner}}]{Kondo_destruction_Si_Pixley}%
  \BibitemOpen
  \bibfield  {author} {\bibinfo {author} {\bibfnamefont {Q.}~\bibnamefont
  {Si}}, \bibinfo {author} {\bibfnamefont {J.~H.}\ \bibnamefont {Pixley}},
  \bibinfo {author} {\bibfnamefont {E.}~\bibnamefont {Nica}}, \bibinfo {author}
  {\bibfnamefont {S.~J.}\ \bibnamefont {Yamamoto}}, \bibinfo {author}
  {\bibfnamefont {P.}~\bibnamefont {Goswami}}, \bibinfo {author} {\bibfnamefont
  {R.}~\bibnamefont {Yu}},\ and\ \bibinfo {author} {\bibfnamefont
  {S.}~\bibnamefont {Kirchner}},\ }\bibfield  {title} {\bibinfo {title} {Kondo
  destruction and quantum criticality in kondo lattice systems},\ }\href
  {https://doi.org/10.7566/JPSJ.83.061005} {\bibfield  {journal} {\bibinfo
  {journal} {Journal of the Physical Society of Japan}\ }\textbf {\bibinfo
  {volume} {83}},\ \bibinfo {pages} {061005} (\bibinfo {year} {2014})},\
  \Eprint {https://arxiv.org/abs/https://doi.org/10.7566/JPSJ.83.061005}
  {https://doi.org/10.7566/JPSJ.83.061005} \BibitemShut {NoStop}%
\bibitem [{\citenamefont {Checkelsky}\ \emph {et~al.}(2024)\citenamefont
  {Checkelsky}, \citenamefont {Bernevig}, \citenamefont {Coleman},
  \citenamefont {Si},\ and\ \citenamefont
  {Paschen}}]{Checkelsky_et_al_flat_bands_Kondo_review}%
  \BibitemOpen
  \bibfield  {author} {\bibinfo {author} {\bibfnamefont {J.~G.}\ \bibnamefont
  {Checkelsky}}, \bibinfo {author} {\bibfnamefont {B.~A.}\ \bibnamefont
  {Bernevig}}, \bibinfo {author} {\bibfnamefont {P.}~\bibnamefont {Coleman}},
  \bibinfo {author} {\bibfnamefont {Q.}~\bibnamefont {Si}},\ and\ \bibinfo
  {author} {\bibfnamefont {S.}~\bibnamefont {Paschen}},\ }\bibfield  {title}
  {\bibinfo {title} {Flat bands, strange metals and the kondo effect},\ }\href
  {https://doi.org/10.1038/s41578-023-00644-z} {\bibfield  {journal} {\bibinfo
  {journal} {Nature Reviews Materials}\ }\textbf {\bibinfo {volume} {9}},\
  \bibinfo {pages} {509} (\bibinfo {year} {2024})}\BibitemShut {NoStop}%
\bibitem [{\citenamefont {Georges}\ and\ \citenamefont
  {Kotliar}(2024)}]{Georges_Kotliar_Hunds_popular}%
  \BibitemOpen
  \bibfield  {author} {\bibinfo {author} {\bibfnamefont {A.}~\bibnamefont
  {Georges}}\ and\ \bibinfo {author} {\bibfnamefont {G.}~\bibnamefont
  {Kotliar}},\ }\bibfield  {title} {\bibinfo {title} {The hund-metal path to
  strong electronic correlations},\ }\href
  {https://doi.org/10.1063/pt.wqrz.qpjx} {\bibfield  {journal} {\bibinfo
  {journal} {Physics Today}\ }\textbf {\bibinfo {volume} {77}},\ \bibinfo
  {pages} {46} (\bibinfo {year} {2024})},\ \Eprint
  {https://arxiv.org/abs/https://pubs.aip.org/physicstoday/article-pdf/77/4/46/20086423/46\_1\_pt.wqrz.qpjx.pdf}
  {https://pubs.aip.org/physicstoday/article-pdf/77/4/46/20086423/46\_1\_pt.wqrz.qpjx.pdf}
  \BibitemShut {NoStop}%
\bibitem [{\citenamefont {Schr{\"o}der}\ \emph {et~al.}(2000)\citenamefont
  {Schr{\"o}der}, \citenamefont {Aeppli}, \citenamefont {Coldea}, \citenamefont
  {Adams}, \citenamefont {Stockert}, \citenamefont {L{\"o}hneysen},
  \citenamefont {Bucher}, \citenamefont {Ramazashvili},\ and\ \citenamefont
  {Coleman}}]{Coleman_AFM_heavy_fermions}%
  \BibitemOpen
  \bibfield  {author} {\bibinfo {author} {\bibfnamefont {A.}~\bibnamefont
  {Schr{\"o}der}}, \bibinfo {author} {\bibfnamefont {G.}~\bibnamefont
  {Aeppli}}, \bibinfo {author} {\bibfnamefont {R.}~\bibnamefont {Coldea}},
  \bibinfo {author} {\bibfnamefont {M.}~\bibnamefont {Adams}}, \bibinfo
  {author} {\bibfnamefont {O.}~\bibnamefont {Stockert}}, \bibinfo {author}
  {\bibfnamefont {H.~v.}\ \bibnamefont {L{\"o}hneysen}}, \bibinfo {author}
  {\bibfnamefont {E.}~\bibnamefont {Bucher}}, \bibinfo {author} {\bibfnamefont
  {R.}~\bibnamefont {Ramazashvili}},\ and\ \bibinfo {author} {\bibfnamefont
  {P.}~\bibnamefont {Coleman}},\ }\bibfield  {title} {\bibinfo {title} {Onset
  of antiferromagnetism in heavy-fermion metals},\ }\href
  {https://doi.org/10.1038/35030039} {\bibfield  {journal} {\bibinfo  {journal}
  {Nature}\ }\textbf {\bibinfo {volume} {407}},\ \bibinfo {pages} {351}
  (\bibinfo {year} {2000})}\BibitemShut {NoStop}%
\bibitem [{\citenamefont {Pixley}\ \emph {et~al.}(2013)\citenamefont {Pixley},
  \citenamefont {Kirchner}, \citenamefont {Ingersent},\ and\ \citenamefont
  {Si}}]{PhysRevB.88.245111}%
  \BibitemOpen
  \bibfield  {author} {\bibinfo {author} {\bibfnamefont {J.~H.}\ \bibnamefont
  {Pixley}}, \bibinfo {author} {\bibfnamefont {S.}~\bibnamefont {Kirchner}},
  \bibinfo {author} {\bibfnamefont {K.}~\bibnamefont {Ingersent}},\ and\
  \bibinfo {author} {\bibfnamefont {Q.}~\bibnamefont {Si}},\ }\bibfield
  {title} {\bibinfo {title} {Quantum criticality in the pseudogap bose-fermi
  anderson and kondo models: Interplay between fermion- and boson-induced kondo
  destruction},\ }\href {https://doi.org/10.1103/PhysRevB.88.245111} {\bibfield
   {journal} {\bibinfo  {journal} {Phys. Rev. B}\ }\textbf {\bibinfo {volume}
  {88}},\ \bibinfo {pages} {245111} (\bibinfo {year} {2013})}\BibitemShut
  {NoStop}%
\bibitem [{\citenamefont {Patri}\ and\ \citenamefont
  {Kim}(2020)}]{PhysRevX.10.041021}%
  \BibitemOpen
  \bibfield  {author} {\bibinfo {author} {\bibfnamefont {A.~S.}\ \bibnamefont
  {Patri}}\ and\ \bibinfo {author} {\bibfnamefont {Y.~B.}\ \bibnamefont
  {Kim}},\ }\bibfield  {title} {\bibinfo {title} {Critical theory of non-fermi
  liquid fixed point in multipolar kondo problem},\ }\href
  {https://doi.org/10.1103/PhysRevX.10.041021} {\bibfield  {journal} {\bibinfo
  {journal} {Phys. Rev. X}\ }\textbf {\bibinfo {volume} {10}},\ \bibinfo
  {pages} {041021} (\bibinfo {year} {2020})}\BibitemShut {NoStop}%
\bibitem [{\citenamefont {Andrei}\ \emph {et~al.}(1983)\citenamefont {Andrei},
  \citenamefont {Furuya},\ and\ \citenamefont
  {Lowenstein}}]{Kondo_problem_RevModPhys}%
  \BibitemOpen
  \bibfield  {author} {\bibinfo {author} {\bibfnamefont {N.}~\bibnamefont
  {Andrei}}, \bibinfo {author} {\bibfnamefont {K.}~\bibnamefont {Furuya}},\
  and\ \bibinfo {author} {\bibfnamefont {J.~H.}\ \bibnamefont {Lowenstein}},\
  }\bibfield  {title} {\bibinfo {title} {Solution of the kondo problem},\
  }\href {https://doi.org/10.1103/RevModPhys.55.331} {\bibfield  {journal}
  {\bibinfo  {journal} {Rev. Mod. Phys.}\ }\textbf {\bibinfo {volume} {55}},\
  \bibinfo {pages} {331} (\bibinfo {year} {1983})}\BibitemShut {NoStop}%
\bibitem [{\citenamefont {Larkin}\ and\ \citenamefont
  {Melnikov}(1972)}]{Larkin_Melnikov}%
  \BibitemOpen
  \bibfield  {author} {\bibinfo {author} {\bibfnamefont {A.~I.}\ \bibnamefont
  {Larkin}}\ and\ \bibinfo {author} {\bibfnamefont {V.~I.}\ \bibnamefont
  {Melnikov}},\ }\bibfield  {title} {\bibinfo {title} {Magnetic impurities in
  an almost magnetic metal},\ }\href@noop {} {\bibfield  {journal} {\bibinfo
  {journal} {Sov. Phys. JETP}\ }\textbf {\bibinfo {volume} {34}},\ \bibinfo
  {pages} {656} (\bibinfo {year} {1972})}\BibitemShut {NoStop}%
\bibitem [{\citenamefont {Maebashi}\ \emph {et~al.}(2002)\citenamefont
  {Maebashi}, \citenamefont {Miyake},\ and\ \citenamefont
  {Varma}}]{Maebashi_Varma_Ferro_2002}%
  \BibitemOpen
  \bibfield  {author} {\bibinfo {author} {\bibfnamefont {H.}~\bibnamefont
  {Maebashi}}, \bibinfo {author} {\bibfnamefont {K.}~\bibnamefont {Miyake}},\
  and\ \bibinfo {author} {\bibfnamefont {C.~M.}\ \bibnamefont {Varma}},\
  }\bibfield  {title} {\bibinfo {title} {Singular effects of impurities near
  the ferromagnetic quantum-critical point},\ }\href
  {https://doi.org/10.1103/PhysRevLett.88.226403} {\bibfield  {journal}
  {\bibinfo  {journal} {Phys. Rev. Lett.}\ }\textbf {\bibinfo {volume} {88}},\
  \bibinfo {pages} {226403} (\bibinfo {year} {2002})}\BibitemShut {NoStop}%
\bibitem [{\citenamefont {Maebashi}\ \emph {et~al.}(2005)\citenamefont
  {Maebashi}, \citenamefont {Miyake},\ and\ \citenamefont
  {Varma}}]{Varma_Kondo_effect_AFM_2005}%
  \BibitemOpen
  \bibfield  {author} {\bibinfo {author} {\bibfnamefont {H.}~\bibnamefont
  {Maebashi}}, \bibinfo {author} {\bibfnamefont {K.}~\bibnamefont {Miyake}},\
  and\ \bibinfo {author} {\bibfnamefont {C.~M.}\ \bibnamefont {Varma}},\
  }\bibfield  {title} {\bibinfo {title} {Undressing the kondo effect near the
  antiferromagnetic quantum critical point},\ }\href
  {https://doi.org/10.1103/PhysRevLett.95.207207} {\bibfield  {journal}
  {\bibinfo  {journal} {Phys. Rev. Lett.}\ }\textbf {\bibinfo {volume} {95}},\
  \bibinfo {pages} {207207} (\bibinfo {year} {2005})}\BibitemShut {NoStop}%
\bibitem [{\citenamefont {Aji}\ \emph {et~al.}(2008)\citenamefont {Aji},
  \citenamefont {Varma},\ and\ \citenamefont
  {Vekhter}}]{Varma_Kondo_effect_AFM_phase_2008}%
  \BibitemOpen
  \bibfield  {author} {\bibinfo {author} {\bibfnamefont {V.}~\bibnamefont
  {Aji}}, \bibinfo {author} {\bibfnamefont {C.~M.}\ \bibnamefont {Varma}},\
  and\ \bibinfo {author} {\bibfnamefont {I.}~\bibnamefont {Vekhter}},\
  }\bibfield  {title} {\bibinfo {title} {Kondo effect in an antiferromagnetic
  metal: Renormalization group analysis and a variational calculation},\ }\href
  {https://doi.org/10.1103/PhysRevB.77.224426} {\bibfield  {journal} {\bibinfo
  {journal} {Phys. Rev. B}\ }\textbf {\bibinfo {volume} {77}},\ \bibinfo
  {pages} {224426} (\bibinfo {year} {2008})}\BibitemShut {NoStop}%
\bibitem [{\citenamefont {Nagaosa}\ and\ \citenamefont
  {Lee}(1997)}]{PhysRevLett.79.3755}%
  \BibitemOpen
  \bibfield  {author} {\bibinfo {author} {\bibfnamefont {N.}~\bibnamefont
  {Nagaosa}}\ and\ \bibinfo {author} {\bibfnamefont {P.~A.}\ \bibnamefont
  {Lee}},\ }\bibfield  {title} {\bibinfo {title} {Kondo effect in high-
  ${T}_{c}$ cuprates},\ }\href {https://doi.org/10.1103/PhysRevLett.79.3755}
  {\bibfield  {journal} {\bibinfo  {journal} {Phys. Rev. Lett.}\ }\textbf
  {\bibinfo {volume} {79}},\ \bibinfo {pages} {3755} (\bibinfo {year}
  {1997})}\BibitemShut {NoStop}%
\bibitem [{\citenamefont {Custers}\ \emph {et~al.}(2003)\citenamefont
  {Custers}, \citenamefont {Gegenwart}, \citenamefont {Wilhelm}, \citenamefont
  {Neumaier}, \citenamefont {Tokiwa}, \citenamefont {Trovarelli}, \citenamefont
  {Geibel}, \citenamefont {Steglich}, \citenamefont {P{\'e}pin},\ and\
  \citenamefont {Coleman}}]{CUSTERS}%
  \BibitemOpen
  \bibfield  {author} {\bibinfo {author} {\bibfnamefont {J.}~\bibnamefont
  {Custers}}, \bibinfo {author} {\bibfnamefont {P.}~\bibnamefont {Gegenwart}},
  \bibinfo {author} {\bibfnamefont {H.}~\bibnamefont {Wilhelm}}, \bibinfo
  {author} {\bibfnamefont {K.}~\bibnamefont {Neumaier}}, \bibinfo {author}
  {\bibfnamefont {Y.}~\bibnamefont {Tokiwa}}, \bibinfo {author} {\bibfnamefont
  {O.}~\bibnamefont {Trovarelli}}, \bibinfo {author} {\bibfnamefont
  {C.}~\bibnamefont {Geibel}}, \bibinfo {author} {\bibfnamefont
  {F.}~\bibnamefont {Steglich}}, \bibinfo {author} {\bibfnamefont
  {C.}~\bibnamefont {P{\'e}pin}},\ and\ \bibinfo {author} {\bibfnamefont
  {P.}~\bibnamefont {Coleman}},\ }\bibfield  {title} {\bibinfo {title} {The
  break-up of heavy electrons at a quantum critical point},\ }\href
  {https://doi.org/10.1038/nature01774} {\bibfield  {journal} {\bibinfo
  {journal} {Nature}\ }\textbf {\bibinfo {volume} {424}},\ \bibinfo {pages}
  {524 EP } (\bibinfo {year} {2003})}\BibitemShut {NoStop}%
\bibitem [{\citenamefont {Lee}(2018)}]{SUNGSIKREVIEW}%
  \BibitemOpen
  \bibfield  {author} {\bibinfo {author} {\bibfnamefont {S.-S.}\ \bibnamefont
  {Lee}},\ }\bibfield  {title} {\bibinfo {title} {Recent developments in
  non-fermi liquid theory},\ }\href
  {https://doi.org/10.1146/annurev-conmatphys-031016-025531} {\bibfield
  {journal} {\bibinfo  {journal} {Annu. Rev. of Condens. Matter Phys.}\
  }\textbf {\bibinfo {volume} {9}},\ \bibinfo {pages} {227} (\bibinfo {year}
  {2018})}\BibitemShut {NoStop}%
\bibitem [{\citenamefont {Abanov}\ \emph {et~al.}(2003)\citenamefont {Abanov},
  \citenamefont {Chubukov},\ and\ \citenamefont {Schmalian}}]{ABANOV3}%
  \BibitemOpen
  \bibfield  {author} {\bibinfo {author} {\bibfnamefont {A.}~\bibnamefont
  {Abanov}}, \bibinfo {author} {\bibfnamefont {A.~V.}\ \bibnamefont
  {Chubukov}},\ and\ \bibinfo {author} {\bibfnamefont {J.}~\bibnamefont
  {Schmalian}},\ }\bibfield  {title} {\bibinfo {title} {Quantum-critical theory
  of the spin-fermion model and its application to cuprates: Normal state
  analysis},\ }\href {https://doi.org/10.1080/0001873021000057123} {\bibfield
  {journal} {\bibinfo  {journal} {Adv. Phys.}\ }\textbf {\bibinfo {volume}
  {52}},\ \bibinfo {pages} {119} (\bibinfo {year} {2003})}\BibitemShut
  {NoStop}%
\bibitem [{\citenamefont {Metlitski}\ and\ \citenamefont
  {Sachdev}(2010)}]{MAX2}%
  \BibitemOpen
  \bibfield  {author} {\bibinfo {author} {\bibfnamefont {M.~A.}\ \bibnamefont
  {Metlitski}}\ and\ \bibinfo {author} {\bibfnamefont {S.}~\bibnamefont
  {Sachdev}},\ }\bibfield  {title} {\bibinfo {title} {Quantum phase transitions
  of metals in two spatial dimensions. ii. spin density wave order},\ }\href
  {https://doi.org/10.1103/PhysRevB.82.075128} {\bibfield  {journal} {\bibinfo
  {journal} {Phys. Rev. B}\ }\textbf {\bibinfo {volume} {82}},\ \bibinfo
  {pages} {075128} (\bibinfo {year} {2010})}\BibitemShut {NoStop}%
\bibitem [{\citenamefont {Schlief}\ \emph {et~al.}(2017)\citenamefont
  {Schlief}, \citenamefont {Lunts},\ and\ \citenamefont {Lee}}]{SCHLIEF}%
  \BibitemOpen
  \bibfield  {author} {\bibinfo {author} {\bibfnamefont {A.}~\bibnamefont
  {Schlief}}, \bibinfo {author} {\bibfnamefont {P.}~\bibnamefont {Lunts}},\
  and\ \bibinfo {author} {\bibfnamefont {S.-S.}\ \bibnamefont {Lee}},\
  }\bibfield  {title} {\bibinfo {title} {Exact critical exponents for the
  antiferromagnetic quantum critical metal in two dimensions},\ }\href
  {https://doi.org/10.1103/PhysRevX.7.021010} {\bibfield  {journal} {\bibinfo
  {journal} {Phys. Rev. X}\ }\textbf {\bibinfo {volume} {7}},\ \bibinfo {pages}
  {021010} (\bibinfo {year} {2017})}\BibitemShut {NoStop}%
\bibitem [{\citenamefont {Borges}\ \emph {et~al.}(2023)\citenamefont {Borges},
  \citenamefont {Borissov}, \citenamefont {Singh}, \citenamefont {Schlief},\
  and\ \citenamefont {Lee}}]{BORGES2023169221}%
  \BibitemOpen
  \bibfield  {author} {\bibinfo {author} {\bibfnamefont {F.}~\bibnamefont
  {Borges}}, \bibinfo {author} {\bibfnamefont {A.}~\bibnamefont {Borissov}},
  \bibinfo {author} {\bibfnamefont {A.}~\bibnamefont {Singh}}, \bibinfo
  {author} {\bibfnamefont {A.}~\bibnamefont {Schlief}},\ and\ \bibinfo {author}
  {\bibfnamefont {S.-S.}\ \bibnamefont {Lee}},\ }\bibfield  {title} {\bibinfo
  {title} {Field-theoretic functional renormalization group formalism for
  non-fermi liquids and its application to the antiferromagnetic quantum
  critical metal in two dimensions},\ }\href
  {https://doi.org/https://doi.org/10.1016/j.aop.2023.169221} {\bibfield
  {journal} {\bibinfo  {journal} {Annals of Physics}\ }\textbf {\bibinfo
  {volume} {450}},\ \bibinfo {pages} {169221} (\bibinfo {year}
  {2023})}\BibitemShut {NoStop}%
\bibitem [{\citenamefont {Wang}\ \emph {et~al.}(2017)\citenamefont {Wang},
  \citenamefont {Schattner}, \citenamefont {Berg},\ and\ \citenamefont
  {Fernandes}}]{PhysRevB.95.174520}%
  \BibitemOpen
  \bibfield  {author} {\bibinfo {author} {\bibfnamefont {X.}~\bibnamefont
  {Wang}}, \bibinfo {author} {\bibfnamefont {Y.}~\bibnamefont {Schattner}},
  \bibinfo {author} {\bibfnamefont {E.}~\bibnamefont {Berg}},\ and\ \bibinfo
  {author} {\bibfnamefont {R.~M.}\ \bibnamefont {Fernandes}},\ }\bibfield
  {title} {\bibinfo {title} {Superconductivity mediated by quantum critical
  antiferromagnetic fluctuations: The rise and fall of hot spots},\ }\href
  {https://doi.org/10.1103/PhysRevB.95.174520} {\bibfield  {journal} {\bibinfo
  {journal} {Phys. Rev. B}\ }\textbf {\bibinfo {volume} {95}},\ \bibinfo
  {pages} {174520} (\bibinfo {year} {2017})}\BibitemShut {NoStop}%
\bibitem [{\citenamefont {Popov}\ and\ \citenamefont
  {Fedotov}(1988)}]{popov1988functional}%
  \BibitemOpen
  \bibfield  {author} {\bibinfo {author} {\bibfnamefont {V.~N.}\ \bibnamefont
  {Popov}}\ and\ \bibinfo {author} {\bibfnamefont {S.}~\bibnamefont
  {Fedotov}},\ }\bibfield  {title} {\bibinfo {title} {The
  functional-integration method and diagram technique for spin systems},\
  }\href@noop {} {\bibfield  {journal} {\bibinfo  {journal} {Zh. Eksp. Teor.
  Fiz}\ }\textbf {\bibinfo {volume} {94}},\ \bibinfo {pages} {183} (\bibinfo
  {year} {1988})}\BibitemShut {NoStop}%
\bibitem [{\citenamefont {Kukreja}\ \emph {et~al.}(2024)\citenamefont
  {Kukreja}, \citenamefont {Besharat},\ and\ \citenamefont
  {Lee}}]{PhysRevB.110.155142}%
  \BibitemOpen
  \bibfield  {author} {\bibinfo {author} {\bibfnamefont {S.}~\bibnamefont
  {Kukreja}}, \bibinfo {author} {\bibfnamefont {A.}~\bibnamefont {Besharat}},\
  and\ \bibinfo {author} {\bibfnamefont {S.-S.}\ \bibnamefont {Lee}},\
  }\bibfield  {title} {\bibinfo {title} {Projective fixed points for non-fermi
  liquids: A case study of the ising-nematic quantum critical metal},\ }\href
  {https://doi.org/10.1103/PhysRevB.110.155142} {\bibfield  {journal} {\bibinfo
   {journal} {Phys. Rev. B}\ }\textbf {\bibinfo {volume} {110}},\ \bibinfo
  {pages} {155142} (\bibinfo {year} {2024})}\BibitemShut {NoStop}%
\bibitem [{\citenamefont {Mandal}\ and\ \citenamefont {Lee}(2015)}]{IPSITA}%
  \BibitemOpen
  \bibfield  {author} {\bibinfo {author} {\bibfnamefont {I.}~\bibnamefont
  {Mandal}}\ and\ \bibinfo {author} {\bibfnamefont {S.-S.}\ \bibnamefont
  {Lee}},\ }\bibfield  {title} {\bibinfo {title} {Ultraviolet/infrared mixing
  in non-fermi liquids},\ }\href {https://doi.org/10.1103/PhysRevB.92.035141}
  {\bibfield  {journal} {\bibinfo  {journal} {Phys. Rev. B}\ }\textbf {\bibinfo
  {volume} {92}},\ \bibinfo {pages} {035141} (\bibinfo {year}
  {2015})}\BibitemShut {NoStop}%
\bibitem [{\citenamefont {Ye}\ \emph {et~al.}(2022)\citenamefont {Ye},
  \citenamefont {Lee},\ and\ \citenamefont {Zou}}]{PhysRevLett.128.106402}%
  \BibitemOpen
  \bibfield  {author} {\bibinfo {author} {\bibfnamefont {W.}~\bibnamefont
  {Ye}}, \bibinfo {author} {\bibfnamefont {S.-S.}\ \bibnamefont {Lee}},\ and\
  \bibinfo {author} {\bibfnamefont {L.}~\bibnamefont {Zou}},\ }\bibfield
  {title} {\bibinfo {title} {Ultraviolet-infrared mixing in marginal fermi
  liquids},\ }\href {https://doi.org/10.1103/PhysRevLett.128.106402} {\bibfield
   {journal} {\bibinfo  {journal} {Phys. Rev. Lett.}\ }\textbf {\bibinfo
  {volume} {128}},\ \bibinfo {pages} {106402} (\bibinfo {year}
  {2022})}\BibitemShut {NoStop}%
\bibitem [{\citenamefont {Zhu}\ and\ \citenamefont
  {Si}(2002)}]{PhysRevB.66.024426}%
  \BibitemOpen
  \bibfield  {author} {\bibinfo {author} {\bibfnamefont {L.}~\bibnamefont
  {Zhu}}\ and\ \bibinfo {author} {\bibfnamefont {Q.}~\bibnamefont {Si}},\
  }\bibfield  {title} {\bibinfo {title} {Critical local-moment fluctuations in
  the bose-fermi kondo model},\ }\href
  {https://doi.org/10.1103/PhysRevB.66.024426} {\bibfield  {journal} {\bibinfo
  {journal} {Phys. Rev. B}\ }\textbf {\bibinfo {volume} {66}},\ \bibinfo
  {pages} {024426} (\bibinfo {year} {2002})}\BibitemShut {NoStop}%
\bibitem [{\citenamefont {Zar\'and}\ and\ \citenamefont
  {Demler}(2002)}]{PhysRevB.66.024427}%
  \BibitemOpen
  \bibfield  {author} {\bibinfo {author} {\bibfnamefont {G.}~\bibnamefont
  {Zar\'and}}\ and\ \bibinfo {author} {\bibfnamefont {E.}~\bibnamefont
  {Demler}},\ }\bibfield  {title} {\bibinfo {title} {Quantum phase transitions
  in the bose-fermi kondo model},\ }\href
  {https://doi.org/10.1103/PhysRevB.66.024427} {\bibfield  {journal} {\bibinfo
  {journal} {Phys. Rev. B}\ }\textbf {\bibinfo {volume} {66}},\ \bibinfo
  {pages} {024427} (\bibinfo {year} {2002})}\BibitemShut {NoStop}%
\bibitem [{\citenamefont {Sengupta}(2000)}]{PhysRevB.61.4041}%
  \BibitemOpen
  \bibfield  {author} {\bibinfo {author} {\bibfnamefont {A.~M.}\ \bibnamefont
  {Sengupta}},\ }\bibfield  {title} {\bibinfo {title} {Spin in a fluctuating
  field: The bose(+fermi) kondo models},\ }\href
  {https://doi.org/10.1103/PhysRevB.61.4041} {\bibfield  {journal} {\bibinfo
  {journal} {Phys. Rev. B}\ }\textbf {\bibinfo {volume} {61}},\ \bibinfo
  {pages} {4041} (\bibinfo {year} {2000})}\BibitemShut {NoStop}%
\bibitem [{\citenamefont {Wilson}(1975)}]{Wilson_RMP}%
  \BibitemOpen
  \bibfield  {author} {\bibinfo {author} {\bibfnamefont {K.~G.}\ \bibnamefont
  {Wilson}},\ }\bibfield  {title} {\bibinfo {title} {The renormalization group:
  Critical phenomena and the kondo problem},\ }\href
  {https://doi.org/10.1103/RevModPhys.47.773} {\bibfield  {journal} {\bibinfo
  {journal} {Rev. Mod. Phys.}\ }\textbf {\bibinfo {volume} {47}},\ \bibinfo
  {pages} {773} (\bibinfo {year} {1975})}\BibitemShut {NoStop}%
\bibitem [{\citenamefont {Andrei}(1980)}]{Andrei_1980}%
  \BibitemOpen
  \bibfield  {author} {\bibinfo {author} {\bibfnamefont {N.}~\bibnamefont
  {Andrei}},\ }\bibfield  {title} {\bibinfo {title} {Diagonalization of the
  kondo hamiltonian},\ }\href {https://doi.org/10.1103/PhysRevLett.45.379}
  {\bibfield  {journal} {\bibinfo  {journal} {Phys. Rev. Lett.}\ }\textbf
  {\bibinfo {volume} {45}},\ \bibinfo {pages} {379} (\bibinfo {year}
  {1980})}\BibitemShut {NoStop}%
\bibitem [{\citenamefont {Andrei}\ and\ \citenamefont
  {Destri}(1984)}]{Andrei_Destri_1984}%
  \BibitemOpen
  \bibfield  {author} {\bibinfo {author} {\bibfnamefont {N.}~\bibnamefont
  {Andrei}}\ and\ \bibinfo {author} {\bibfnamefont {C.}~\bibnamefont
  {Destri}},\ }\bibfield  {title} {\bibinfo {title} {Solution of the
  multichannel kondo problem},\ }\href
  {https://doi.org/10.1103/PhysRevLett.52.364} {\bibfield  {journal} {\bibinfo
  {journal} {Phys. Rev. Lett.}\ }\textbf {\bibinfo {volume} {52}},\ \bibinfo
  {pages} {364} (\bibinfo {year} {1984})}\BibitemShut {NoStop}%
\bibitem [{\citenamefont {Affleck}(1990)}]{AFFLECK1990517}%
  \BibitemOpen
  \bibfield  {author} {\bibinfo {author} {\bibfnamefont {I.}~\bibnamefont
  {Affleck}},\ }\bibfield  {title} {\bibinfo {title} {A current algebra
  approach to the kondo effect},\ }\href
  {https://doi.org/https://doi.org/10.1016/0550-3213(90)90440-O} {\bibfield
  {journal} {\bibinfo  {journal} {Nuclear Physics B}\ }\textbf {\bibinfo
  {volume} {336}},\ \bibinfo {pages} {517} (\bibinfo {year}
  {1990})}\BibitemShut {NoStop}%
\bibitem [{\citenamefont {Affleck}\ and\ \citenamefont
  {Ludwig}(1991)}]{AFFLECK1991849}%
  \BibitemOpen
  \bibfield  {author} {\bibinfo {author} {\bibfnamefont {I.}~\bibnamefont
  {Affleck}}\ and\ \bibinfo {author} {\bibfnamefont {A.~W.}\ \bibnamefont
  {Ludwig}},\ }\bibfield  {title} {\bibinfo {title} {The kondo effect,
  conformal field theory and fusion rules},\ }\href
  {https://doi.org/https://doi.org/10.1016/0550-3213(91)90109-B} {\bibfield
  {journal} {\bibinfo  {journal} {Nuclear Physics B}\ }\textbf {\bibinfo
  {volume} {352}},\ \bibinfo {pages} {849} (\bibinfo {year}
  {1991})}\BibitemShut {NoStop}%
\end{thebibliography}%

\newpage
\onecolumngrid
\renewcommand\appendixpagename{\centering\large End Matter}
\appendix
\appendixpage

\section{Field-theoretic functional RG scheme
}
\label{sec:vertexfunction}

\begin{figure}[t]
    \centering
    \begin{subfigure}[t]{0.25\linewidth}
        \centering
        \includegraphics[width=\linewidth]{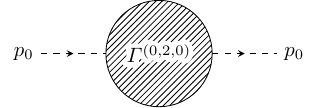}
        \caption{\label{fig:gamma020}}
    \end{subfigure}
    \begin{subfigure}[t]{0.25\linewidth}
        \centering
        \includegraphics[width=\linewidth]{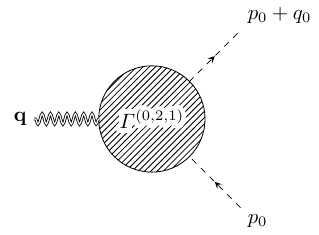}
        \caption{\label{fig:gamma021}}
    \end{subfigure}
    \begin{subfigure}[t]{0.25\linewidth}
        \centering
        \includegraphics[width=\linewidth]{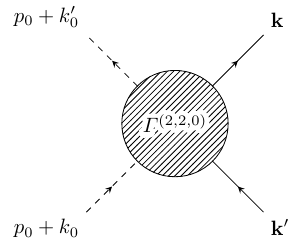}
        \caption{\label{fig:gamma220}}
    \end{subfigure}
    \caption{
({$\color{blue}a$}) The inverse propagator of the peudo-fermion.
(${\color{blue}b}$) The boson-impurity vertex.
(${\color{blue}c}$) The electron-impurity vertex.
Here, the double wiggly lines, the dashed lines,
and the solid lines represent  
the boson propagators, the pseudo-fermion propagators and the electron propagators, respectively.
}
    \label{fig:vertexfuncs}
\end{figure}

In this appendix, we describe the functional renormalization group (RG) scheme, closely following
Ref. \cite{BORGES2023169221}.
Let $\varGamma^{(2m,2n,l)}$ be the vertex function for $2m$ itinerant electrons, $2n$ pseudo fermions and $l$ bosons.
The normalization of the pseudo-fermion field,
and the Kondo coupling function
and the impurity-boson coupling 
are defined through the vertex function so that
    (see \fig{fig:vertexfuncs})
   \begin{align}
    \left.-i\frac{\partial}{\partial p_0}\varGamma^{(0,2,0)}(p_0)\right|_{p_0 = \mu} = & 1 + \mathscr{F}_0,\label{eq:impSEvf}
    \\ \left.\varGamma^{(0,2,1)}(p_0,q_0)\right|_{p_0=q_0/2=\mu} = & \frac{1}{2}\left(g_f + \mathscr{F}_1 \right),\label{eq:impbosonvf}
    \\ \left.\varGamma^{(2,2,0);(N,N')}({\bf k},{\bf k}',p_0)\right|_{
    \begin{array}{l}\scriptsize
    p_0/2 = k'_0 = -k_0 = \mu \\ \scriptsize
    e_N[\vec{k};v^{(N)}_{k}] =
    e_{N'}[\vec{k'};v^{(N')}_{k'}] =0
 \end{array}
    } = & \frac{1}{4\mu}\left[J^{(N,N')}_{k_N,k'_{N'}} + \mathscr{F}^{(N,N')}_{3;k_N,k'_{N'}}\right].
    \label{eq:kondovf}
\end{align}
Here, $\mathscr{F}_0, \mathscr{F}_1$ are scheme-dependent terms that are regular in the $\mu \to 0$ limit. 
For the Kondo coupling,
which is a function of momentum along the Fermi surface, 
not only $\mathscr{F}^{(N,N')}_{3;k_N,k'_{N'}}$ is required to be regular in $\mu$ at each external momentum 
but also its integral done along the Fermi surface with the measure $\frac{dk}{\mu}$ should be regular \cite{BORGES2023169221}.
This stricter condition is necessary to capture an IR divergence that may arise 
for a momentum-integrated dimensionless coupling such as the one in the s-wave Kondo coupling $\sim \int \frac{dk_N}{\mu} 
 J_{k_N,k'_{N'}}$
while $J_{k_N,k'_{N'}}$ at each momentum is regular.

To impose these RG conditions, a local counter term action is added to as \eq{eq:ImpurityAction},
\begin{equation}
\begin{aligned}
&S_1^{\mathrm{C.T.}}
= 
\int\frac{\dd p_0}{2\pi}A_{f,1} i p_0 f^{\dagger}_{\alpha} (p_0)f_{\alpha}(p_0)\\
& +\frac{1}{\mu}
\sum_{N,N'} \int\dd{\bf k}\dd{\bf q}\frac{\dd p_0}{2\pi}A_{f,2}^{(N,N')}(k_N,q_{N'})J^{(N,N')}_{k_N,q_{N'}}\psi^{\dagger}_{N,\sigma}({\bf k})\frac{\vec{\tau}_{\sigma,\sigma'}}{2}\psi_{N',\sigma'}({\bf q})
f^{\dagger}_{\alpha}(p_0+q_0)\frac{\vec{\tau}_{\alpha,\beta}}{2}f_\beta(p_0+k_0)\\
& + g_f\sum_{\alpha,\beta = \uparrow, \downarrow}\int\dd{\bf q}\frac{\dd p_0}{2\pi}A_{f,3} f^{\dagger}_{\alpha}(p_0+q_0)\frac{\Phi_{\alpha,\beta}({\bf q})}{2}f_{\beta}(p_0),
\end{aligned}
\label{eq:CTAction}
\end{equation}
where 
$\Phi_{\alpha,\beta} \equiv 
\vec \phi \cdot
\vec{\tau}_{\alpha,\beta}$.
For the RG conditions of the coupling functions that appear in Eq. \eqref{eq:cleanaction} and their counter terms, we refer the reader to Ref. \cite{BORGES2023169221}.
Adding \eq{eq:CTAction} to \eq{eq:ImpurityAction}, we obtain the renormalized action for the impurity,
\begin{equation}
\begin{aligned}
&S^{\mathrm{Ren}}
= 
\int\frac{\dd p^\mathsf{B}_0}{2\pi}i p^\mathsf{B}_0 f^{\mathsf{B}\dagger}_{\alpha} (p^\mathsf{B}_0)f^\mathsf{B}_{\alpha}(p^\mathsf{B}_0)\\
& +
\sum_{N,N'}
\int\dd{\bf k}^\mathsf{B}\dd{\bf q}^\mathsf{B}\frac{\dd p^\mathsf{B}_0}{2\pi}J^{\mathsf{B}(N,N')}_{k^\mathsf{B}_N,q^\mathsf{B}_N}\psi^{\mathsf{B}\dagger}_{N,\sigma}({\bf k}^\mathsf{B})\frac{\vec{\tau}_{\sigma,\sigma'}}{2}\psi^\mathsf{B}_{N',\sigma'}({\bf q}^\mathsf{B})\cdot f^{\mathsf{B}\dagger}_{\alpha}(p^\mathsf{B}_0+q^\mathsf{B}_0)\frac{\vec{\tau}_{\alpha,\beta}}{2}f^\mathsf{B}_\beta(p^\mathsf{B}_0+k^\mathsf{B}_0)\\
& + \sum_{\alpha,\beta = \uparrow, \downarrow}\int\dd{\bf q}^\mathsf{B}\frac{\dd p^\mathsf{B}_0}{2\pi}g^\mathsf{B}_f f^{\mathsf{B}\dagger}_{\alpha}(p^\mathsf{B}_0+q^\mathsf{B}_0)\frac{\Phi^\mathsf{B}_{\alpha,\beta}({\bf q}^\mathsf{B})}{2}f^\mathsf{B}_{\beta}(p^\mathsf{B}_0).
\end{aligned}
\label{eq:RenAction}
\end{equation}
Here,
\begin{equation}
    \begin{gathered}
k^\mathsf{B}_0 = Z_\tau k_0 , \quad 
\vec{k}^\mathsf{B} = \vec{k}, \quad
\psi^{\mathsf{B}}_{N,\sigma}({\bf k}^\mathsf{B}) = 
\sqrt{Z^{(\psi,N)}(k_N)}\psi_{N,\sigma}({\bf k}), \quad 
\Phi^\mathsf{B}_{\sigma\sigma'}({\bf q}^\mathsf{B}) = \sqrt{Z^{(\Phi)}}\Phi_{\sigma\sigma'}({\bf q}), \quad 
f_\alpha^\mathsf{B}(p^\mathsf{B}_0) = \sqrt{Z^{(f)}}f_\alpha(p_0),\\
%
%
J^{\mathsf{B}(N,N')}_{k^\mathsf{B}_N,q^\mathsf{B}_{N'}} =  \frac{Z_{f,2}^{(N,N')}(k_N,q_{N'})}{ Z_\tau^3 Z^{(f)}  \sqrt{Z^{(\psi,N)}(k_N)Z^{(\psi,N')}(q_{N'})} } \mu^{-1}J^{(N,N')}_{k_N,q_{N'}}, \quad 
g^\mathsf{B}_f =  \frac{Z_{f,3}}{Z_\tau^2 Z^{(f)} \sqrt{Z^{(\Phi)}}}g_f
    \end{gathered}
    \label{eq:barequantities}
\end{equation}
are bare frequency, fields and couplings expressed in terms of the renormalized ones.
$Z_\tau$ is the dynamical critical exponent.
$\sqrt{Z^{(\psi,N)}(k_N)}$
and
$\sqrt{Z^{(\Phi)}}$
are renormalization of the fermion and boson fields, respectively\cite{BORGES2023169221}.
$\quad Z^{(f)} = \frac{Z_{f,1}}{Z_\tau^2}$ is the field renormalization of the pseudo-fermion.
The renormalization factors are given by 
$Z_{f,1} = 1 + A_{f,1}$, 
$Z^{(N,N')}_{f,2}(k_N,q_{N'}) = 1 + A^{(N,N')}_{f,2}(k_N,q_{N'})$ 
and $Z_{f,3} = 1 + A_{f,3}$. 
The beta functions capture how the renormalized couplings, which represent the vertex functions at energy scale $\mu$, runs as $\mu$ is lowered for fixed bare couplings,
\begin{align}
    \beta^{(g_f)} = & g_f\left(2(z-1) + 2\eta^{(f)} + 
    \eta^{(\Phi)}
    -\frac{\dd\log Z_{f,3}}{\dd\log\mu}\right),\label{eq:betagfdefapp}
    \\ \beta^{(J);(N,N')}_{k,k'} = & J^{(N,N')}_{k,k'}\left(1 + 3(z-1)+2\eta^{(f)}
    +\eta^{(\psi,N)}_{k}+\eta^{(\psi,N')}_{k'}
    -\frac{\dd \log Z^{(N,N')}_{f,2}(k,k')}{\dd\log\mu}\right),\label{eq:betaKondodefapp}
\end{align}
where
\begin{equation}
    \begin{aligned}
        z = 1 +\frac{\dd \log Z_\tau}{\dd \log \mu},& \quad 
        \eta^{(\psi,N)}_{k} = \frac{1}{2}\frac{\dd \log Z^{(\psi,N)}(k)}{\dd \log\mu}, & \quad 
        \eta^{(\Phi)} = \frac{1}{2}\frac{\dd \log Z^{(\Phi)}}{\dd \log\mu}, \quad & 
        \eta^{(f)} = \frac{1}{2}\frac{\dd \log Z^{(f)}}{\dd \log\mu}.
    \end{aligned}
\end{equation}
In the small $v_k$ limit, the dynamical critical exponent and the anomalous dimensions are given by
\cite{SCHLIEF,BORGES2023169221}
\begin{align}
    z =  1 + \frac{3}{4\pi^2}w,  ~~~~~
    \eta^{(\Phi)} =  \frac{1}{4\pi} w \log \frac{1}{w},
   ~~~~~ 
     \eta^{(\psi,N)}_{k_N} =  \frac{3 (g^{(N)}_{k_N,k_N})^2
    }{4\pi^2c V^{(\overline{N})}_{F,k_N}}\frac{\mu}{\mu + 
    2 cv^{(N)}_{k_N}
    \abs{k_N}} - (z-1),
\end{align}
where $w \equiv v_0/c(v_0)$
with
$c(v_0) = \sqrt{\frac{v_0}{16}\log\left(\frac{1}{v_0}\right)}$.

\section{Quantum corrections}
\label{sec:quantumcorrections}

\begin{figure}[t]
    \centering
    \begin{subfigure}[t]{0.49\linewidth}
        \centering
        \includegraphics{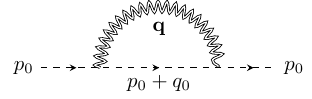}
        \caption{\label{fig:ImpSE}}
    \end{subfigure}
    \begin{subfigure}[t]{0.49\linewidth}
        \centering
        \includegraphics{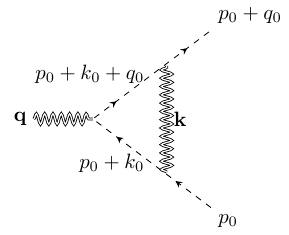}
        \caption{\label{fig:gfVC}}
    \end{subfigure}
    \centering
    \begin{subfigure}[t]{0.32\linewidth}
        \centering
        \includegraphics[scale=0.7]{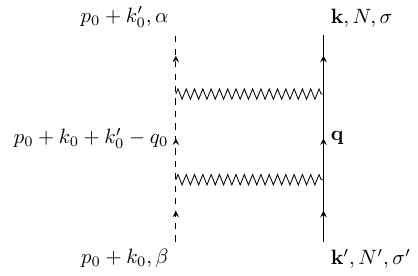}
        \caption{\label{fig:kondoc}}
    \end{subfigure}
    \begin{subfigure}[t]{0.32\linewidth}
        \centering
        \includegraphics[scale=0.7]{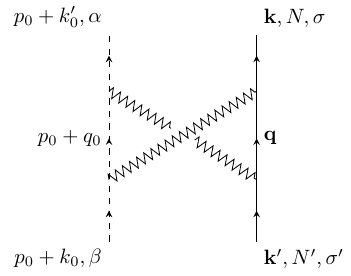}
        \caption{\label{fig:kondod}}
    \end{subfigure}
    \begin{subfigure}[t]{0.32\linewidth}
        \centering
        \includegraphics[scale=0.7]{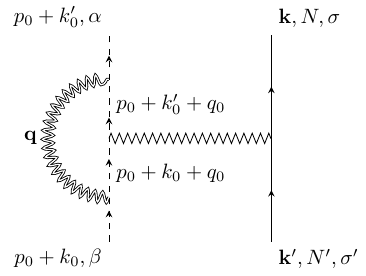}
        \caption{\label{fig:kondoi}}
    \end{subfigure}
\caption{
The one-loop diagrams that renormalize the pseudo-fermion, the boson-impurity coupling and the Kondo coupling.
(${\color{blue}a}$) The psuedo-fermion self energy. 
(${\color{blue}b}$) The	boson-impurity vertex correction. 
(${\color{blue}c}$-${\color{blue}e}$) The vertex correction for the Kondo coupling.
The boson propagator is non-perturbatively dressed by particle-hole excitations.
}
\label{fig:KondoDiagrams}
\end{figure}

In this appendix, we present the one-loop quantum corrections that renormalize the pseudo-fermion self-energy, the boson-impurity coupling and the electron-impurity (Kondo) coupling. 
The diagrams that are relevant to the boson-impurity coupling and the Kondo coupling are shown in  \fig{fig:KondoDiagrams}.
For the quantum corrections for the impurity-free theory, we refer the readers to Refs. \cite{
SCHLIEF,BORGES2023169221}.

\subsection{Pseudo-fermion self-energy}
\label{ssec:impSE}

The one-loop self-energy 
in \fig{fig:ImpSE} gives
\begin{equation}
    {\bf \Sigma}^{(f);\mathrm{1L}}(p_0) = -\frac{3 g_f^2}{4}\int\dd{\bf q}D({\bf q})G_f(p_0 + q_0)
    = -\frac{3 g_f^2}{4}\int\frac{\dd q_0}{2\pi}\bar{D}(q_0)G_f(p_0 + q_0),
    \label{eq:impSEapp}
\end{equation}
where
\begin{equation}
    \bar{D}(q_0) = \int \frac{\dd q_x \dd q_y}{(2\pi)^2}D({\bf q}) = \frac{1}{\pi^2 c^2}\left\lbrace 2 c \Lambda_b\log\left(1 + \frac{c \Lambda_b}{c \Lambda_b + \abs{q_0}}\right) + \abs{q_0}\left[\log\left(1 + \frac{c \Lambda_b}{c \Lambda_b + \abs{q_0}}\right) - \log\left(1 + \frac{c \Lambda_b}{\abs{q_0}}\right)\right]\right\rbrace
    \label{eq:bosonpropintmom}
\end{equation}
is the local boson propagator at the impurity site.
Here, 
$\Lambda_b$ is the momentum cutoff below which the self-energy of the boson generated from the particle-hole excitation
$\Pi({\bf q}) = \left[ \abs{q_0} + c(\abs{q_x}+\abs{q_y})\right]^{-1}$  becomes dominant over the bare boson propagator dropped in this calculation.
The logarithmic UV divergence $\log \Lambda_b/q_0$ in the coefficient of $|q_0|$ arises because the bosons with large momenta up to $\Lambda_b$ have a significant spectral weight at low energies\cite{SCHLIEF}.
While the precise value of $\Lambda_b$ is not important,  it is crucial that $\Lambda_b$ is a fixed UV momentum cutoff.
From now on, we set $\Lambda_b = \Lambda/c$,
where $\Lambda$ is an energy cutoff.
The frequency integration $q_0$ results in
\begin{equation}
    {\bf \Sigma}^{(f);\mathrm{1L}}(p_0) \approx i p_0\frac{ 3 g_f^2\log ^2\left(\frac{\Lambda}{\abs{p_0} }\right)}{8 \pi ^3 c^2}-i p_0 \frac{3 g_f^2\log (2) \log \left(\frac{\Lambda}{\abs{p_0} }\right)}{4\pi ^3 c^2}+i p_0\frac{3 g_f^2 \log \left(\frac{\Lambda}{\abs{p_0} }\right)}{4\pi ^3 c^2}
\end{equation}
to the leading order in small $p_0$. With the renormalization conditions \eqref{eq:impSEvf}, the counter term in the small $\mu$ limit is chosen to be
\begin{equation}
    A_{f,1}^{\mathrm{1L}} = \frac{3 g_f^2 \log (2) \log \left(\frac{\Lambda}{\mu }\right)}{4 \pi ^3 c^2}-\frac{3 g_f^2 \log ^2\left(\frac{\Lambda}{\mu }\right)}{8 \pi ^3 c^2}.\label{eq:Af1result}
\end{equation}

\subsection{Boson-impurity vertex correction}
\label{ssec:gfQC}

The one-loop boson-impurity vertex correction in Fig. \ref{fig:gfVC} reads
\begin{equation}
    {\bf \Gamma}^{\mathrm{1L}}(p_0,q_0) = -\frac{g_f^3}{8}\int\dd {\bf k} D({\bf k})G_f(p_0+k_0)G_f(p_0+q_0+k_0)
    = -\frac{g_f^3}{8}\int \frac{\dd k_0}{2\pi} \bar{D}(k_0)G_f(p_0+k_0)G_f(p_0+q_0+k_0).
    \label{eq:gfvertexintegral}
\end{equation}
Integrating $k_0$ results in the quantum correction at external frequency $\mu$,
\begin{equation}
    \left.{\bf \Gamma}^{\mathrm{1L}}(p_0,q_0)\right|_{p_0=q_0/2=\mu} 
    =  -\frac{g_f^3 \log ^2\left(\frac{c \Lambda_b}{\mu }\right)}{16 \pi ^3 c^2}+\frac{g_f^3(\log(108)-2) \log \left(\frac{c \Lambda_b}{\mu }\right)}{16 \pi ^3 c^2}
\end{equation}
up to terms that are regular in $\mu$.
The corresponding counter term is
\begin{equation}
    A_{f,3}^{\mathrm{1L}} = \frac{g_f^2 \log ^2\left(\frac{c \Lambda_b}{\mu }\right)}{8 \pi ^3 c^2}-\frac{g_f^2(\log(108)-2) \log \left(\frac{c \Lambda_b}{\mu }\right)}{8 \pi ^3 c^2}.
    \label{eqb7}
\end{equation}

\subsection{Kondo coupling vertex corrections}
\label{ssec:kondoQC}

The diagrams in Figs. \ref{fig:kondoc} and \ref{fig:kondod}, originally computed by Kondo, give
\begin{align}
    {\bf \Gamma}_{FL(PP);(k_N,k'_{N'},p_0)}^{(N,N');\spmqty{\alpha & \sigma \\ \beta & \sigma'};(1L)} = & \left(3\delta_{\sigma\sigma'}\delta_{\alpha\beta}-2\vec{\tau}_{\sigma\sigma'}\cdot\vec{\tau}_{\alpha\beta}\right){\bf \Gamma}_{FL(PP);(k_{N},k'_{N'},p_0)}^{(N,N');(1L)},
    \label{eq:gammaCKshort}\\
    {\bf \Gamma}_{FL(PH);(k_N,k'_{N'},p_0)}^{(N,N');\spmqty{\alpha & \sigma \\ \beta & \sigma'};(1L)} = & \left(3\delta_{\sigma\sigma'}\delta_{\alpha\beta}+2\vec{\tau}_{\sigma\sigma'}\cdot\vec{\tau}_{\alpha\beta}\right){\bf \Gamma}_{FL(PH);(k_{N},k'_{N'},p_0)}^{(N,N');(1L)},
    \label{eq:gammaDKshort}
\end{align}
where
\begin{align}
    {\bf \Gamma}_{FL(PP);(k_{N},k'_{N'},p_0)}^{(N,N');(1L)} = & -
   \sum_M 
   \int\dd {\bf q}\frac{J^{(N,M)}_{k_N,q_M}J^{(M,N')}_{q_M,k'_{N'}}}{16\mu^2}G_M({\bf q})G_f(p_0+k_0+k'_0-q_0),
    \label{eq:gammaCkondointegral}\\
    {\bf \Gamma}_{FL(PH);(k_{N},k'_{N'},p_0)}^{(N,N');(1L)} = & -\sum_M
    \int\dd {\bf q}\frac{J^{(N,M)}_{k_N,q_M}J^{(M,N')}_{q_M,k'_{N'}}}{16\mu^2} G_M({\bf q})G_f(p_0 + q_0).
    \label{eq:gammaDkondointegral}
\end{align}
%
%
At the external frequencies, $k'_0 = - k_0 = p_0/2 = \mu$ set by the RG condition, \eqref{eq:gammaCkondointegral} gives rise to
\begin{equation}
 {\bf \Gamma}_{FL(PP);(k_{N},k'_{N'})}^{(N,N');(1L)} = 
- \sum_M 
\int\frac{\dd q_M}{2\pi}\frac{J^{(N,M)}_{k_N,q_M}J^{(M,N')}_{q_M,k'_{N'}}}{32\pi\mu^2 V^{(M)}_{F,q_M}}\log\left(\frac{\Lambda}{\mu}\right).
\end{equation}
The $q_M$ integral is left undone because we do not know the explicit form of the Kondo coupling in prior.
The corresponding counter term becomes
$ {\bf \Gamma}_{CT;FL(PP);(k_{N},k'_{N'})}^{(N,N');(1L)} = \sum_M
\int\frac{\dd q_M}{2\pi}\frac{J^{(N,M)}_{k_N,q_M}J^{(M,N')}_{q_M,k'_{N'}}}{32\pi\mu^2 V^{(M)}_{F,q_M}}\log\left(\frac{\Lambda}{\mu}\right)$.
The evaluation of Eq. \eqref{eq:gammaDkondointegral} is similar and the resulting counter term is given by
$ {\bf \Gamma}_{CT;FL(PH);(k_{N},k'_{N'})}^{(N,N');(1L)} $
$= - {\bf \Gamma}_{CT;FL(PP);(k_{N},k'_{N'})}^{(N,N');(1L)} $.
In the net counter term, 
the density-density interaction cancels, leading to 
\begin{equation}
    {\bf \Gamma}_{CT;FL;(k_N,k_{N'})}^{(N,N');\spmqty{\alpha & \sigma \\ \beta & \sigma'};(1L)} = -\vec{\tau}_{\sigma\sigma'}\cdot\vec{\tau}_{\alpha\beta}
    \sum_M
    \int\frac{\dd q_M}{2\pi}\frac{J^{(N,M)}_{k_N,q_M}J^{(M,N')}_{q_M,k'_{N'}}}{8\pi\mu^2 V^{(M)}_{F,q_M}}\log\left(\frac{\Lambda}{\mu}\right).
\end{equation}
It contributes to the IR beta functional as
\begin{equation}
    \left. 4\mu\frac{\partial{\bf \Gamma}_{CT;FL;(k_N,k_{N'})}^{(N,N');\spmqty{\alpha & \sigma \\ \beta & \sigma'};(1L)}}{\partial\log\mu} \right|_{J_B}
    = \vec{\tau}_{\sigma\sigma'}\cdot\vec{\tau}_{\alpha\beta}
    \sum_M
    \int\frac{\dd q_M}{2\pi \mu V^{(M)}_{F,q_M}}\frac{J^{(N,M)}_{k_N,q_M}J^{(M,N')}_{q_M,k'_{N'}}}{2\pi},
    \label{eq:countertermKondoCD}
\end{equation}
where the derivative with respect to $\log \mu$ is done with fixed bare Kondo coupling, which is
$
J^{B(N,M)}_{k_N,q_M}= 
J^{(N,M)}_{k_N,q_M}/\mu$ 
to the leading order.

In AFQCM, 
\fig{fig:kondoi} gives rise to an additional vertex correction to Kondo coupling.
Since the critical boson renormalizes the Kondo coupling in the exact same way as it renormalizes the boson-impurity vertex through
\eqref{eq:gfvertexintegral},
the multiplicative renormalization factor is identical to that of 
    \eq{eqb7},

\begin{equation}
    {\bf \Gamma}_{CT;NFL}^{(N,N');(1L)} =  
    \frac{
    J^{(N,N')}_{k_N,k'_{N'}}}{4\mu}
    \left[
    \frac{ g_f^2 \log^2\left(\frac{\Lambda}{\mu}\right)}{8\pi^3c^2}
    -
    \frac{
    g_f^2
    \left(\log(108) - 2\right)\log\left(\frac{\Lambda}{\mu}\right)}{8\pi^3 c^2}
    \right]. 
\end{equation}
For the beta function, the term that is proportional to $\log^2 \Lambda/\mu$ dominates and gives
\begin{equation}
     4\mu\frac{\partial {\bf \Gamma}_{CT;NFL}^{(N,N');\spmqty{\alpha & \sigma \\ \beta & \sigma'};(1L)}}{\partial\log\mu} =  -\vec{\tau}_{\sigma\sigma'}\cdot\vec{\tau}_{\alpha\beta}\frac{g_f^2  J^{(N,N')}_{k_N,k'_{N'}}}{4 \pi ^3 c^2}\log \left(\frac{\Lambda}{\mu }\right).
     \label{eq:countertermKondoI}
\end{equation}

\section{
The solution of the beta functions
}
\label{sec:qptkondo}

\begin{figure}
    \centering
    \includegraphics[width=0.5\linewidth]{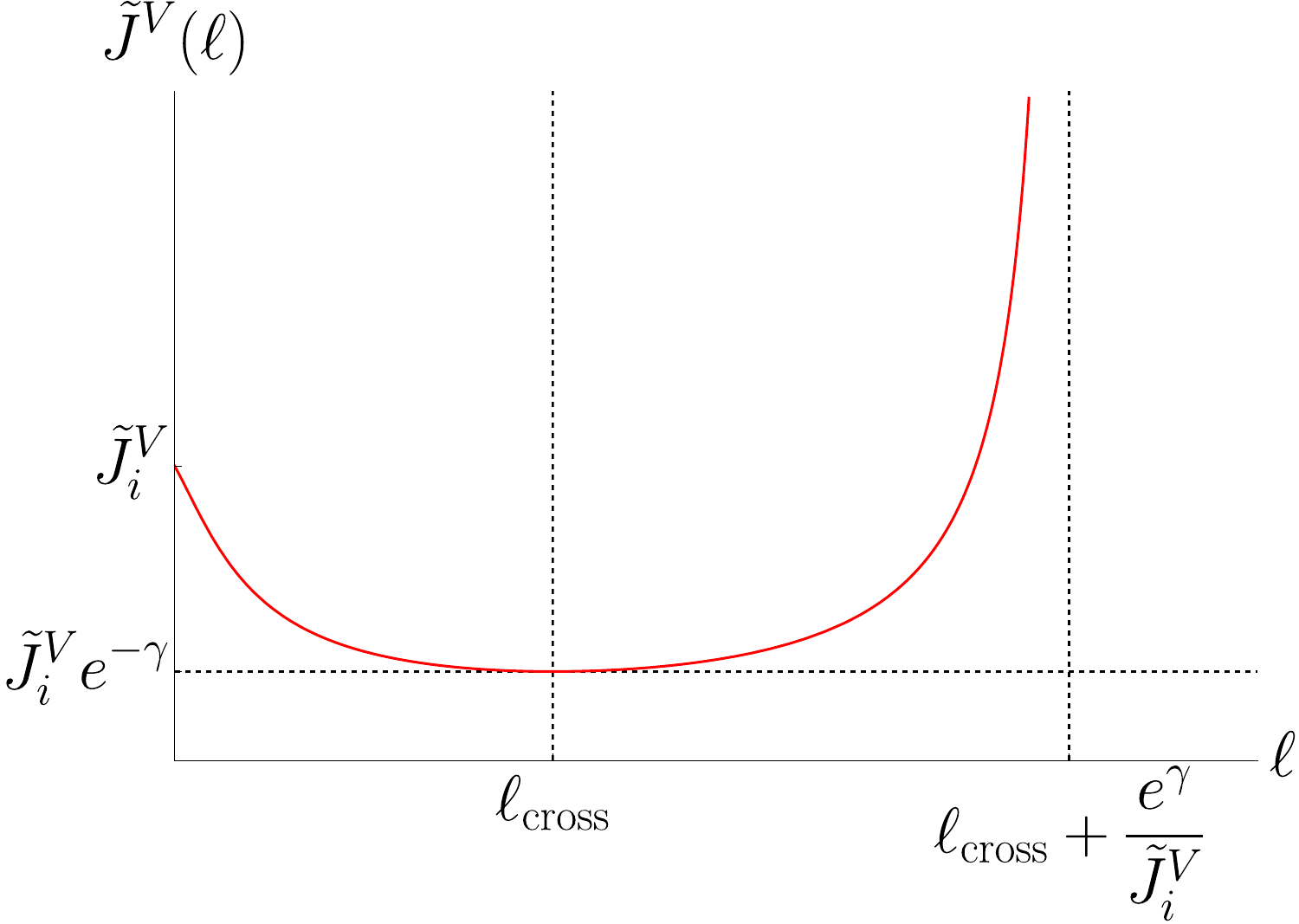}
    \caption{
A schematic RG flow of Kondo coupling $\tJV$ in the small $\tJV_i$ limit.
At short distances 
below a crossover scale 
$\ell_{\mathrm{cross}}$,
$\tJV$ is suppressed by the anomalous dimension $\eta_f$.
Beyond the crossover scale,
the anomalous dimension becomes negligible,
and $\tJV$ is enhanced as in  Fermi liquids.
Therefore, the scale at which $\tJV$ becomes $O(1)$ is 
given by $\ell_K \sim
\ell_{\mathrm{cross}}
+ \frac{1}{\tJV(
\ell_{\mathrm{cross}}
)}
$.
 }
    \label{fig:Jcross}
\end{figure}
In this appendix, we compute Kondo scale $\ell_K$ as a function of the bare parameters
$\tJV_i$, $g_{f,i}$ and $\ell_0$
defined at the short-distance cutoff scale $\ell_i$
by solving Eqs. \eqref{eq:betaJ}  and \eqref{eq:betag}.
In the small $\tJV_i$ limit,
the flow of $\tJV$ can be understood in two steps.
At short distance scales smaller than a crossover scale denoted by $\ell_{\mathrm{cross}}$,
$\tJV$ undergoes a suppression due to the anomalous dimension generated from the boson-impurity coupling.
Its flow is described by 
$\frac{\partial \tilde J^V(\ell)}{\partial \ell} =  
-\eta_{f}(\ell) \tilde{J}^V(\ell) 
$.
Then, $\tJV$ at the crossover scale becomes
$\tJV(\ell_{\mathrm{cross}})
\sim \tJV_i e^{-\gamma}$ with
$\gamma = \int_{\ell_i}^{ 
\ell_{\mathrm{cross}}
} d\ell' \eta_f(\ell')$.
For $\ell \gg \ell_{\mathrm{cross}}$,
the flow of $\tJV$ can be approximated by 
$\frac{\partial \tilde J^V(\ell)}{\partial \ell} = \left(\tilde{J}^V(\ell)\right)^2$
as is in Fermi liquids,
and
$\tJV$ becomes strong at Kondo scale 
$\ell_K \sim 
\ell_{\mathrm{cross}}
+ e^\gamma/\tJV_i$.
This is illustrated in     \fig{fig:Jcross}.
Since $\ell_{\mathrm{cross}}$ is largely independent of $\tJV_i$ 
in the small $\tJV_i$ limit,
$\ell_K$ is mainly determined by the renormalized Kondo coupling 
at $\ell_{\mathrm{cross}}$.

\begin{figure}[t]
    \centering
    \begin{subfigure}[t]{0.3\linewidth}
        \centering
        \includegraphics[width=\linewidth]{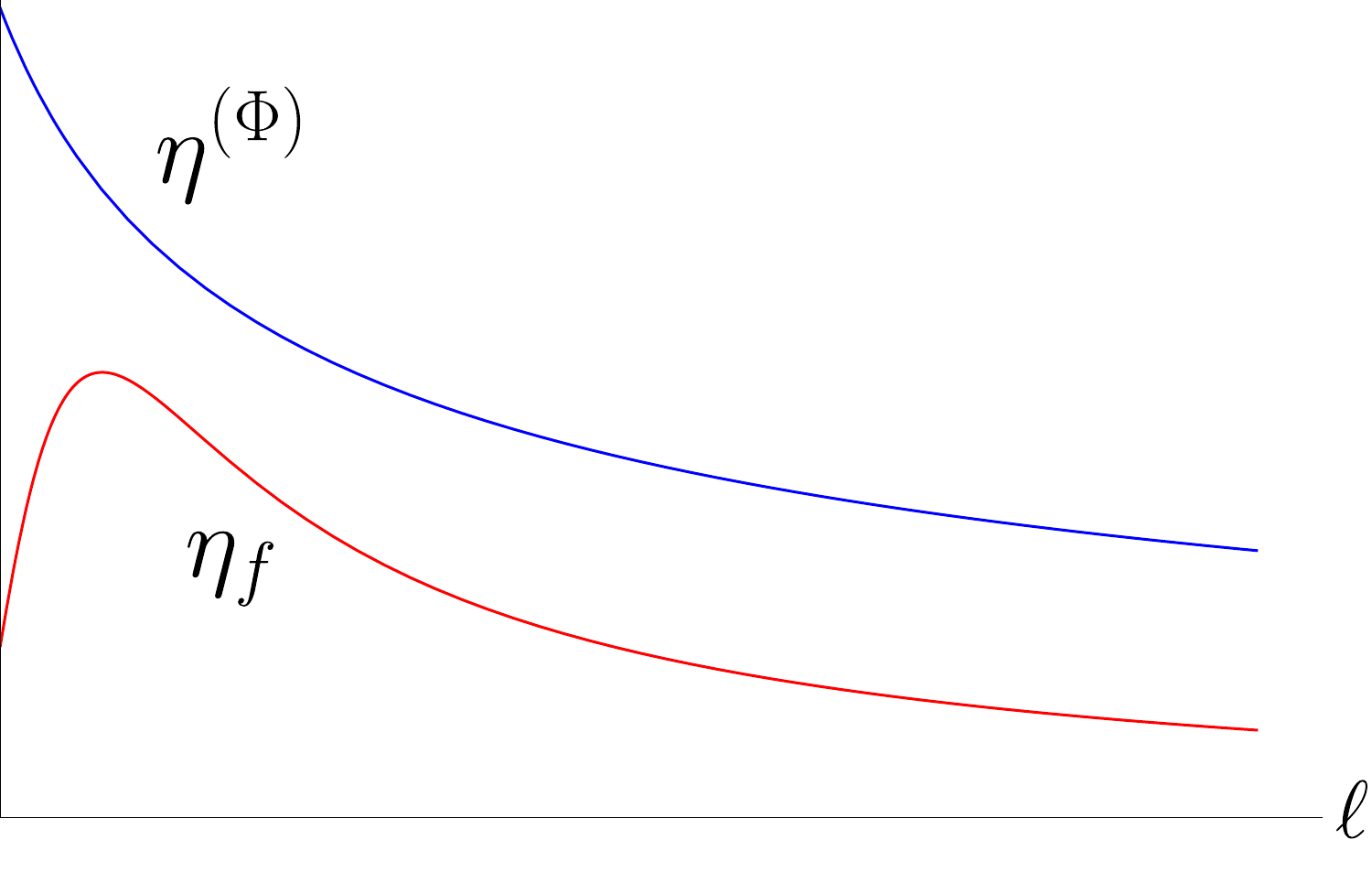}
        \caption{\label{fig:example1}}
    \end{subfigure}\hfill
    \begin{subfigure}[t]{0.3\linewidth}
        \centering
        \includegraphics[width=\linewidth]{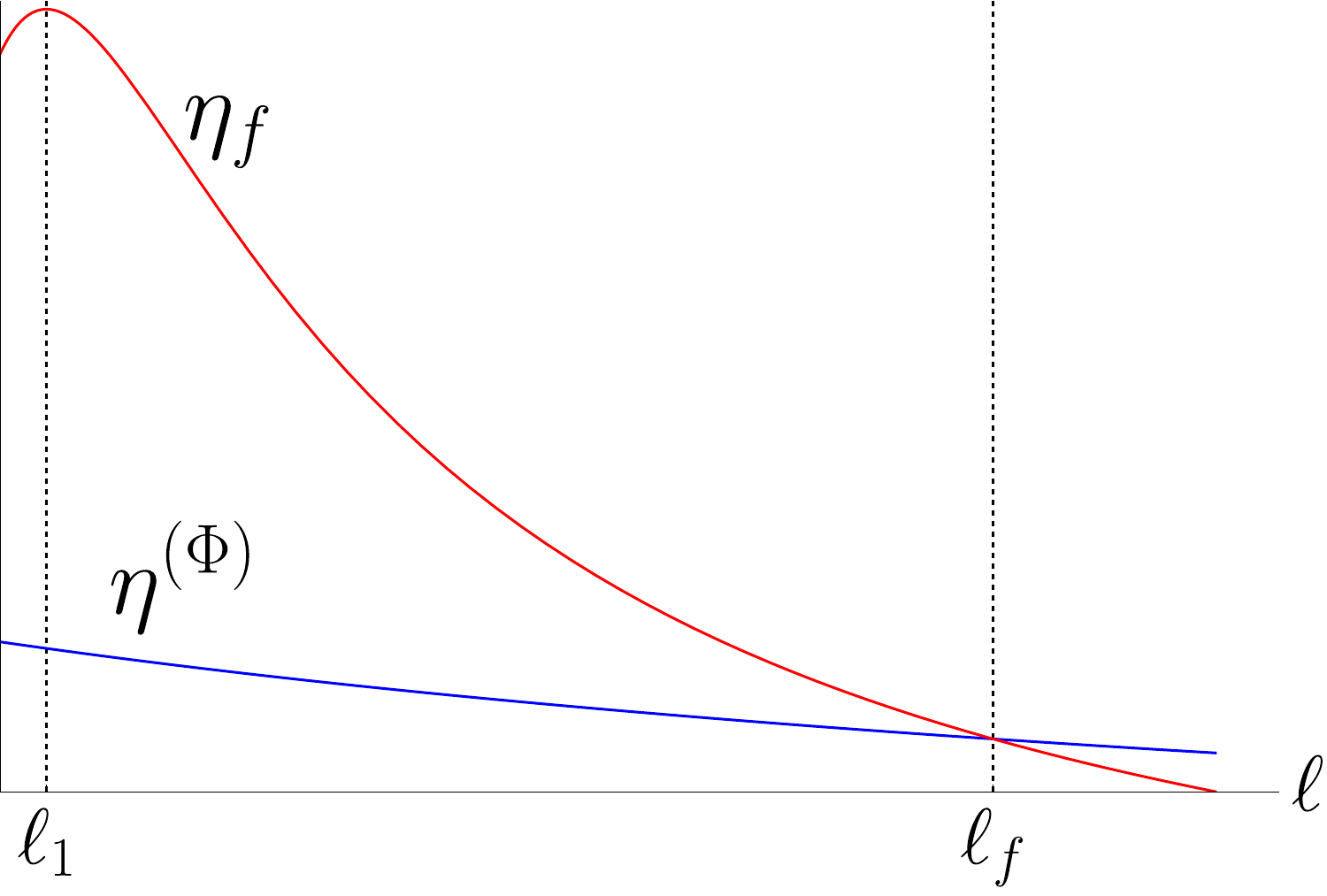}
        \caption{\label{fig:example2}}
    \end{subfigure}\hfill
    \begin{subfigure}[t]{0.3\linewidth}
        \centering
        \includegraphics[width=\linewidth]{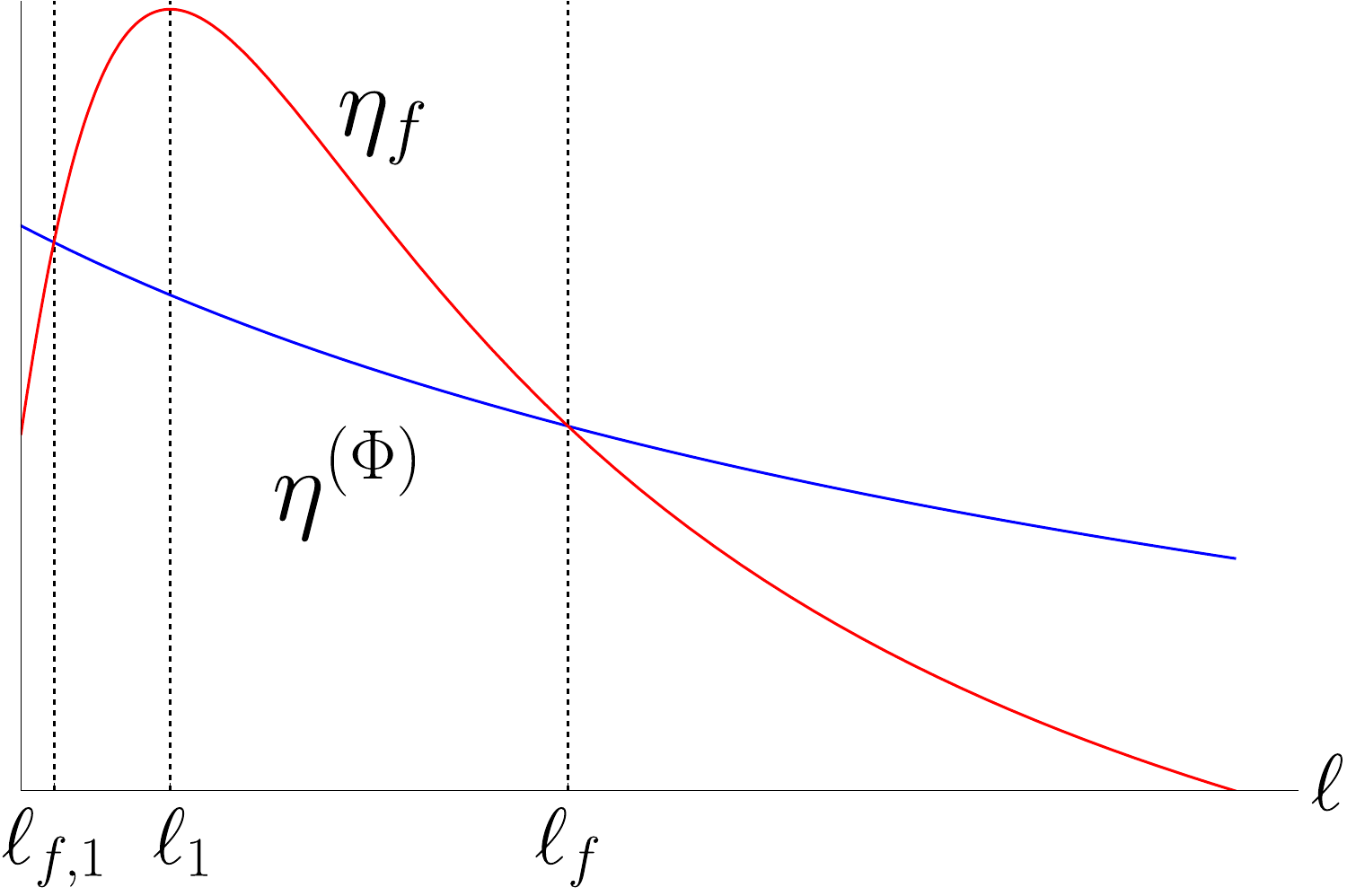}
        \caption{\label{fig:example3}}
    \end{subfigure}
    \caption{
The renormalization group flow of the boson-impurity coupling ($g_f$) is controlled by  $\etaphi(\ell)$, the correction to the anomalous dimension of the boson at non-zero nesting angle,
and
$\eta_f(\ell) = \tgf(\ell) \ell/\pi^3$, the anomalous dimension generated from $g_f$ itself.
$g_f$ exhibits different behaviors, depending on the relative magnitudes of  $\etaphi_i \sim 1/\sqrt{\ell_0}$ and $\sqrt{\tgfi}$.
(${\color{blue} a}$) 
In this case,
$\tgfi$ is small enough
 that $\eta_f$ remains negligible compared to $\etaphi$ at all scales
(Appendix C \ref{app:kondoscale1}).
(${\color{blue}b}$)
In this case,
$\tgf$ is dominant over $\etaphi$ from the UV scale all the way to a crossover scale $\ell_f \sim \sqrt{\ell_0}$
(Appendix C  \ref{app:kondoscale2} a).
(${\color{blue}c}$) 
This is similar to case (${\color{blue} b}$) except that there is an additional window of scale $\ell_i < \ell < \ell_{f,1}$ in which $\etaphi$ is larger than $\eta_f$
(Appendix C  \ref{app:kondoscale2} b).
}
    \label{fig:etas}
\end{figure}

In order to compute this suppression factor of the Kondo coupling ($e^{-\gamma}$)
at $\ell_{\mathrm{cross}}$,
we first need to know how $\eta_f$ evolves as a function of $\ell$ by solving the beta function of $g_f$.
The flow of $g_f$ exhibits different behaviors depending on the relative magnitude between $1/\ell_0$ and $\tgf=g_f^2/c^2$ at $\ell_i$.
Three different cases are illustrated in \fig{fig:etas}.
Below, we discuss each case one by one.

\subsection{
$\tgfi \ll 1/\ell_0$ 
}
\label{app:kondoscale1}

In this case, the boson-impurity coupling is weak,
and its flow is mainly controlled by the anomalous dimension of the boson $\etaphi$.
In particular,
$\tgf$ can be dropped in \eq{eq:betag} in the beta function of $g_f$, 
$\frac{\partial g_f(\ell) }{\partial \ell} = -\frac{1}{4\pi}w\log\left(\frac{1}{w}\right)g_f(\ell)$,
where $w=v/c$.
The finite $w$-correction to the anomalous dimension of the boson makes the boson-impurity coupling irrelevant and flow to zero at large distance scales.
From $w(\ell)$ determined from \eq{eq:c}, one obtains
$g_f(\ell) =  e^{-\frac{\sqrt{\ell +\ell_0}-\sqrt{\ell_0+\ell_i}}{\sqrt{3}}}g_{f,i}$,
and the effective boson-impurity coupling $\tgf = g_f^2/c^2$ becomes
\begin{equation}
    \tgf(\ell) = \frac{\ell_0 +\ell }{\ell_0+\ell_i}
    e^{-\frac{2 \left(\sqrt{\ell+\ell_0}-\sqrt{\ell_i +\ell_0}\right)}{\sqrt{3}}}
    \tilde g_{f,i},
\label{eq:gtildelargel03}
\end{equation}
where subscript $i$ denote the coupling defined at initial length scale $\ell_i \sim O(1)$.
As $\ell$ increases
$\eta_f = (\tgf \ell / \pi^3)$
reaches its maximum 
$\tgfi \sez$
around scale $\sez$
before it decays exponentially.
Since $\etaphi$ only decays as $1/\sqrt{\ell+\ell_0}$, $\etaphi$ remains dominant over $\eta_f$ at all scales. 
Therefore, \eq{eq:gtildelargel03} is valid at all $\ell$.

We can now solve \eq{eq5}
using \eq{eq:gtildelargel03}.
If the Kondo coupling is weak at $\ell_i$, 
$- \eta_f \tJV$ 
dominates over $(\tJV)^2$ at short distance scales.
This allows us to use $ \frac{\partial \tilde J^V(\ell)}{\partial \ell} = 
    -\eta_{f}(\ell)
    \tilde{J}^V(\ell)$
to find $\tJV(\ell)$ before $(\tJV)^2$ becomes dominant:
$\tilde{J}^V(\ell) 
= \tilde{J}^V_i e^{-\gamma_1(\ell)}$
with
\begin{equation}
\gamma_1(\ell) =
\tilde{g}_{f,i}
    \frac{
    \left\lbrace
     \begin{gathered}
         \left[4 \sqrt{3} (\ell_0-15) (\ell +\ell_0)^{3/2}-4 \sqrt{3} (\ell +\ell_0)^{5/2}-30 (\ell +\ell_0)^2\right.  \\ \left. +18 (\ell_0-15) (\ell +\ell_0)+18 \sqrt{3} (\ell_0-15) \sqrt{\ell +\ell_0}+27 (\ell_0-15)\right] e^{\frac{2 \left(\sqrt{\ell_0+\ell_i}-\sqrt{\ell +\ell_0}\right)}{\sqrt{3}}} \\ 
         -4 \sqrt{3} (\ell_0-15) (\ell_0+\ell_i)^{3/2}+4 \sqrt{3} (\ell_0+\ell_i)^{5/2}+30 (\ell_0+\ell_i)^2-18 (\ell_0-15) (\ell_0+\ell_i) \\ -18 \sqrt{3} (\ell_0-15) \sqrt{\ell_0+\ell_i}-27 (\ell_0-15)  \end{gathered} \right\rbrace}{4 \pi ^3 (\ell_0+\ell_i)}.
\label{eq:tildeJVlinsolb}
\end{equation}
At short distance scales,
Kondo coupling is exponentially suppressed due to $\tgf$. 
For $\ell > \sqrt{\ell_0}$, however, $\tgf$ becomes exponentially small, slowing down the decay of $\tJV$.
Therefore, a crossover from the $\eta_f \tJV$-dominated flow to the $(\tJV)^2$-dominated flow occurs at the crossover scale 
$\ell_{\mathrm{cross},1}$ 
which is determined by
$\eta_f(\ell_{\mathrm{cross},1})= \tilde{J}^V_{I}(\ell_{\mathrm{cross},1})$.
We note that 
$\ell_{\mathrm{cross},1} > \sqrt{\ell_0}$ in the small $\tJV_i$ limit. 
At scale greater than $\sqrt{\ell_0}$, 
\eq{eq:tildeJVlinsolb}
saturates to
\begin{equation}
 \tilde{J}^V_{I}
    (\ell \gg \sqrt{\ell_0} ) =
    \tilde{J}^V_i e^{
    - \frac{
   3 \tilde{g}_{f,i}
   \ell_0}{\pi^3}},
\label{eq:tildeJVlinsolasympb}
\end{equation}
and 
$\ell_{\mathrm{cross},1}$ satisfies
$\sqrt{\ell_{\mathrm{cross},1}+\ell_0}-\sqrt{\ell_0}-\frac{\sqrt{3}}{2}\log\left(\frac{(\ell_{\mathrm{cross},1}+\ell_0)\ell_{\mathrm{cross},1}}{\ell_0}\right)  =\frac{\sqrt{3}}{2}\log\left(\frac{\tilde{g}_{f,i}}{\pi^3}\right) + \frac{3^{\frac{3}{2}}\tilde{g}_{f,i}\ell_0}{2\pi^3} + \frac{\sqrt{3}}{2}\log\left(\frac{1}{\tilde{J}^V_i}\right)$.
To the leading in $\tJV$, the crossover scale is given by
\begin{equation}
    \ell_{\mathrm{cross},1} = 
        \frac{3 \sqrt{3} \tilde{g}_{f,i} \ell_0^{3/2}}{\pi ^3}+\sqrt{3} \sqrt{\ell_0} \log \left(\frac{\tilde{g}_{f,i}}{\pi ^3 \tilde{J}^V_i}\right)+\frac{3 \tilde{g}_{f,i}^2 \left(\frac{\pi ^3 \log \left(\frac{\tilde{g}_{f,i}}{\pi ^3 \tilde{J}^V_i}\right)}{\tilde{g}_{f,i}}+3 \ell_0\right)^2}{4 \pi ^6}.
    \label{eq:crossover1}
\end{equation}
For $\ell \gg \ell_{\mathrm{cross},1}$, the flow of the Kondo coupling is governed by
$ \frac{\partial \tilde J^V(\ell)}{\partial \ell} =  
\left(\tilde J^V(\ell)\right)^2$,
and its solution is 
$  \tilde{J}^V_{I}(\ell \geq \ell_{\mathrm{cross},1}) = \frac{1}{\left(\tilde{J}^V_{I}(\ell_{\mathrm{cross},1})\right)^{-1} - (\ell-\ell_{\mathrm{cross},1})}$.
This results in the Kondo scale,
\begin{equation}
    \ell_K = 
\ell_{\mathrm{cross},1} 
+ 
    \frac{e^{\frac{3\tilde{g}_{f,i}\ell_0}{\pi^3}}}{\tilde{J}^V_i}.
\end{equation}
In the small $\tJV$ limit, $\ell_K$ is given by
$\frac{e^{\frac{3\tilde{g}_{f,i}\ell_0}{\pi^3}}}{\tilde{J}^V_i}$ to the leading order.
Compared to the Fermi liquid with the same electronic density of states,
$\ell_K$ is larger by factor of
$e^{\frac{3\tilde{g}_{f,i}\ell_0}{\pi^3}}$.

\subsection{
$1/\ell_0 \ll \tgfi$ 
}
\label{app:kondoscale2}

In this case, the boson-impurity coupling plays the dominant role, 
and the behavior of Kondo coupling exhibits a strong departure from that of Fermi liquids.
Below, we present the solution of the beta function for two sub-cases separately:
(a) $1/\sez \ll \tgfi$, 
(b) $1/\ell_0 \ll \tgfi \ll 1/\sez$.

\subsubsection{$1/\sez \ll \tgfi$} 

In this case, 
$\tgf$ is dominant  so that one can ignore $\etaphi$ in the beta function of $g_f$ at short distance scale.
$g_f$ obeys
$ \frac{\partial g_f(\ell) }{\partial \ell} = - \frac{g_f^3(\ell)\ell }{\pi ^3 c^2}$
with its solution
$g_f(\ell) = \pm\frac{\pi^{5/2} \abs{g_{f,i}}}{\sqrt{\pi^5 + 16 g_{f,i}^2 \left(2\ell^3 + 3\ell^2\ell_0-\ell_i^2(3\ell_0+2\ell_i)\right)}}$.
Then, $\tgf$ becomes
\begin{equation}
\tilde{g}_f(\ell) = \frac{3\pi^3\tilde{g}_{f,i}(\ell+\ell_0)}{3\pi^3 (\ell_0+\ell_i)+\tilde{g}_{f,i}  (2\ell^3+3\ell^2\ell_0-\ell_i^2(3\ell_0+2\ell_i))}.
\label{eq:gfetazerosol}
\end{equation}
While 
$\tgf(\ell) \approx \tilde g_{f,i}$
at short distance scales, it takes a universal form,
$\tgf(\ell) \approx 
\frac{3(\ell+\ell_0)\pi^3}{2\ell^3+3\ell^2\ell_0}$
at long distance scales.
This crossover occurs around
\begin{equation}
    \ell_1 \approx \sqrt{\frac{\pi^3}{\tilde{g}_{f,i}}} \ll \ell_0^{1/4}.
    \label{eq:ell1large}
\end{equation}
According to 
\eq{eq:gfetazerosol},
$\tgf \ell$ decays faster than $\etaphi$ with increasing $\ell$.
Therefore, there must be a crossover scale $\ell_f$ above which $\tgf \ell$ becomes smaller than $\etaphi$.
Equating $\eta^{(\Phi)}(\ell_f)$ and $\tgf \ell_f/\pi^3$ gives $\ell_f^2 - 2\sqrt{3\ell_0}\ell_f + \ell_1^2 =0$ whose only positive solution is 
\begin{equation}
\ell_f = \sqrt{3\ell_0} + \sqrt{3\ell_0 - \ell_1^2}
\approx  2\sqrt{3\ell_0}.
    \label{eq:gfcrossover}
\end{equation}
For $\ell_0 \gg 1$,
we have a hierarchy 
 $\ell_1 \ll \ell_f \ll \ell_0$.
%
For $\ell \gg \ell_f$, the beta function for $g_f$ is given by
    $\frac{\partial g_f(\ell) }{\partial \ell} = -\frac{1}{4\pi}w\log\left(\frac{1}{w}\right)g_f(\ell)$
whose solution is given by
$g_f(\ell \geq \ell_f) =  e^{-\frac{\sqrt{\ell +\ell_0}-\sqrt{\ell_0+\ell_f}}{\sqrt{3}}}g_f(\ell_f)$.
This gives 
$\tgf(\ell)$ that decays exponentially,
\begin{equation}
    \tgf(\ell \geq \ell_f) = \frac{\pi^3 (\ell_0 +\ell) }{\ell_f^2 (\ell_0+\ell_f)}e^{-\frac{2 \left(\sqrt{\ell +\ell_0}-\sqrt{\ell_0+\ell_f}\right)}{\sqrt{3}}}.
    \label{eq:gtildelargel02}
\end{equation}
With $\tgf$, we can now understand how Kondo scale $\ell_K$ depends on $\tJV_i$.
As $\tJV_i$ decreases, $\ell_K$ continuously increases such that it passes three crossover scales $\ell_f$.
Therefore, we consider the following two limiting cases:
$i$) $\ell_K  \ll \ell_f$, 
and
$ii$) $\ell_f \ll \ell_K$. 

\paragraph{$\ell_K \ll \ell_f$ }

In this case,
the flow of $g_f$ is dominated by $\eta_f$ 
at all scales up to $\ell_K$.
Using the large $\ell_0$ expression of Eq. \eqref{eq:gfetazerosol},
$\tgf(\ell) \approx 
\frac{\tilde{g}_{f,i}\pi^3}{\pi^3 + \tilde{g}_{f,i}(\ell^2 - \ell_i^2)}$,
we obtain the solution for 
\eq{eq5} is
\begin{equation}
\tilde{J}^V
(\ell) = \frac{\sqrt{\frac{\pi ^3}{\tilde{g}_{f,i}}} \tilde{J}^V_i}{\sqrt{\frac{\pi ^3}{\tilde{g}_{f,i}}+\ell ^2-\ell_i^2} \left(\sqrt{\frac{\pi ^3}{\tilde{g}_{f,i}}} \tilde{J}^V_i \log \left(\frac{\sqrt{\frac{\pi ^3}{\tilde{g}_{f,i}}}+\ell_i}{\ell + \sqrt{\frac{\pi ^3}{\tilde{g}_{f,i}}+\ell^2 -\ell_i^2}}\right)+1\right)}.
\label{eq:JVtildesol}
\end{equation}
With $\ell_i \sim O(1)$, the Kondo length scale is given by 
\begin{equation}
    \ell_K = \sqrt{\frac{\pi ^3 }{\tilde{g}_{f,i}}}\sinh 
    \left(
  \sqrt{\frac{\tilde{g}_{f,i}}{\pi ^3 }}\frac{1}{ \tilde{J}^V_i}
    \right).
    \label{eq:ellKI}
\end{equation}
For $\tJV_i/\sqrt{\tilde g_{f,i}} \gg 1$,
the bare Kondo coupling is relatively strong that the boson-impurity coupling plays only a minimal role,
and
$\ell_K$ is essentially reduced to that of Fermi liquids with a small correction,
    $\ell_K \approx 
    \frac{1}{\tilde{J}^V_i}
    \left(
    1 + \frac{\tilde{g}_{f,i}}{6 \pi^3 \left(\tilde{J}^V_i\right)^2}
    \right)$.
%
For $\tJV_i/\sqrt{\tilde g_{f,i}} \ll 1$, on the other hand,
the boson-impurity coupling significantly suppresses Kondo coupling,
and $\ell_K$
becomes much larger than that of Fermi liquids,
    $\ell_K \approx 
    \frac{1}{2}\sqrt{\frac{\pi^3}{\tilde{g}_{f,i}}}e^{\sqrt{\frac{\tilde{g}_{f,i}}{\pi^3}}\frac{1}{\tilde{J}^V_i}}$.
Remarkably, the logarithmic Kondo scale $\ell_K$ grows exponentially as $\tJV_i$ decreases.

\paragraph{ $ \ell_f \ll \ell_K$ } \label{app:kondoscale2b}


In the small $\tJV_i$ limit,
$\ell_K$ becomes greater than $\ell_f$.
In this case, one has to solve the beta function for the Kondo coupling in multiple steps.
At short distance scales, the beta function can be approximated as
$\frac{\partial \tilde J^V(\ell)}{\partial \ell} = 
    -\frac{\tgf(\ell) \ell}{\pi^3}
    \tilde{J}^V(\ell)$.
Its solution is
$\tilde{J}^V(\ell) 
= \tilde{J}^V_i e^{-\gamma_2(\ell)}$
with
\begin{equation}
\gamma_2(\ell)=
\frac{\left\lbrace
        \begin{gathered}
        \left[4 \sqrt{3} (\ell_0-15) (\ell +\ell_0)^{3/2}-4 \sqrt{3} (\ell +\ell_0)^{5/2} 
        -30 (\ell +\ell_0)^2+18 (\ell_0-15) (\ell +\ell_0)
        +18 \sqrt{3} (\ell_0-15) \sqrt{\ell +\ell_0} \right.
        \\ \left. +27 (\ell_0-15)\right] e^{\frac{2 \left(\sqrt{\ell_0+\ell_f}-\sqrt{\ell +\ell_0}\right)}{\sqrt{3}}}-4 \sqrt{3} (\ell_0-15) (\ell_0+\ell_f)^{3/2}
        +4 \sqrt{3} (\ell_0+\ell_f)^{5/2}+12 \ell_0 (\ell_0+\ell_f)\\
        +30 \ell_f (\ell_0+\ell_f)+270 (\ell_0+\ell_f)
        -18 \sqrt{3} (\ell_0-15) \sqrt{\ell_0+\ell_f}-27 \ell_0+405 \end{gathered}\right\rbrace}{4 \ell_f^2 (\ell_0+\ell_f)}.
        \label{eq:tildeJVlinsol}
    \end{equation}
This $\tJV$-linear beta function is valid up to a crossover scale 
$\ell_{\mathrm{cross},2}$ above which $(\tJV)^2$ term dominates. 
At the crossover scale,
\eq{eq:tildeJVlinsol} saturates to
\begin{equation}
        \tilde{J}^V
        (\ell_{\mathrm{cross},2}) = e^{-\frac{3}{4}}\frac{ \tilde{J}^V_i}{2}\sqrt{\frac{\pi^3}{3\tilde{g}_{f,i} \ell_0}}
\label{eq:JVIsaturation2}
    \end{equation}
for sufficiently small $\tJV_i$.
From $\eta_f(\ell_{\mathrm{cross},2})= 
\tilde{J}^V_{IIb}(\ell_{\mathrm{cross},2})$, we obtain
the crossover scale,
\begin{equation}
    \begin{aligned}
        \ell_{\mathrm{cross},2} = & -\frac{3}{4} \left(\log \left(\frac{\sqrt{\frac{3\tilde{g}_{f,i}\ell_0}{\pi ^3}}}{6 \ell_0 (\ell_0+2\sqrt{3\ell_0})\tilde{J}^V_i}\right)+\frac{3}{4}\right)^2+\sqrt{3} \sqrt{\ell_0+2\sqrt{3\ell_0}} \log \left(\frac{\sqrt{\frac{3\tilde{g}_{f,i}\ell_0}{\pi ^3}}}{6 \ell_0 (\ell_0+2\sqrt{3\ell_0})\tilde{J}^V_i}\right)
        \\ & +\frac{3}{4} \sqrt{3} \sqrt{\ell_0+2\sqrt{3\ell_0}}+2\sqrt{3\ell_0}.
    \end{aligned}
        \label{eq:ellcf}
    \end{equation}
For $\ell \gg \ell_{\mathrm{cross},2}$,
the Kondo coupling runs as in Fermi liquids,
$\tilde{J}^V_{IIb}(\ell \geq \ell_{\mathrm{cross},2}) = \frac{1}{\left(\tilde{J}^V_{IIa}(\ell_{\mathrm{cross},2})\right)^{-1} - 
(\ell-\ell_{\mathrm{cross},2})}$,
and
the Kondo scale is given by
    \begin{equation}
        \begin{aligned}
            \ell_K = & 
\ell_{\mathrm{cross},2}
+
\frac{2e^{\frac{3}{4}}}{ \tilde{J}^V_i}\sqrt{\frac{3\tilde{g}_{f,i} \ell_0}{\pi^3}}.
        \end{aligned}
        \label{eq:ellKcross2}
    \end{equation}
 To the leading order in $1/\tJV_i$, the Kondo scale becomes
    $\ell_K \sim 
   \frac{1}{\tJV_i} 
   \frac{g_{f,i}}{v_{0,i} \log 1/v_{0,i}}$, where $v_{0,i} \sim 1/(\ell_0\log\ell_0)$.
   For fixed $\tJV_i$ and $g_{f,i}$,
   the Kondo length scale diverges as the bare nesting angle $v_{0,i}$ approaches zero.
The $v_{0,i}$ and $g_{f,i}$ dependences of $\ell_K$ are confirmed through the numerical solution of the beta functions, as is shown in   \fig{fig:phasediagram}.
There is a discrepancy in the prefactor $A$ 
for $\ell_K  =  \frac{A}{\tJV_i}  \frac{g_{f,i}}{v_{0,i} \log 1/v_{0,i}}$
between the analytic estimation and the numerical solution: 
$A_\mathrm{analytic} =  \frac{8 e^{3/4}}{\sqrt{\pi }}$
vs.
$A_\mathrm{numeric} \approx  \frac{8 e^{3/4}}{3\sqrt{\pi }}$.
This is because 
\eq{eq:ellKcross2}
does not accurately account for the $\ell$-dependent anomalous dimension around the crossovers from
the $\eta_f$-dominated flow
to the 
the $\etaphi$-dominated flow.
Since we only keep only one dominant term between $\etaphi$ and $\eta_f$ for the flow of $g_f$,
our analytic result overestimates $g_f$.
This, in turn, makes $\eta_f$ larger, causing an underestimation of 
$\tJV$,
and hence
$\ell_{K,\mathrm{analytic}} > \ell_{K,\mathrm{numeric}}$.
However, this does not affect how $\ell_K$ depends on $g_{f,i}$ and $v_{0,i}$.

\subsubsection{
$1/\ell_0 \ll \tgfi \ll 1/\sez$ 
}
\label{app:kondoscale3}

In this case, $\etaphi_i$ is larger than $\tgfi$ at the UV cut off scale $\ell_i$, but is not large enough to stay dominant over $\tgf$ at all scales.
Since $\eta_f(\ell)$ initially grows as a function of $\ell$, there is a crossover $\ell_{f,1}$, where the term $\tgf \ell/\pi^3$ in the beta function of $g_f$ dominates over $\etaphi$ at crossover scale set by
$\eta^{(\Phi)}(\ell_{f,1}) = \eta_f(\ell_{f,1})$.
The crossover scale is given by
\begin{equation}
    \ell_{f,1} \approx \frac{ \pi ^3}{2\sqrt{3\ell_0} \tilde{g}_{f,i}}.
    \label{eq:newcrossover1}
\end{equation}
At this crossover scale, the Kondo coupling 
becomes
$\tJV_i e^{-\frac{\pi ^3}{24 \tilde{g}_{f,i} \ell_0}}$ 
in the small $\tJV_i$ limit.
The flow in $\ell > \ell_{f,1}$ is then identical to the previous case
discussed in Sec. \ref{app:kondoscale2}.
%
If $\ell_K \ll \ell_f$, $\tJV(\ell)$ is given by
\eq{eq:JVtildesol}
with the replacement of
$\ell_i \to \ell_{f,1}$ 
and
$\tJV_i \to \tJV_i e^{-\frac{\pi ^3}{24 \tilde{g}_{f,i} \ell_0}}$.
This gives the Kondo scale that is analogous to Eq. \eqref{eq:ellKI},
\begin{equation}
    \ell_K = 
    \frac{ \pi ^3}{2\sqrt{3\ell_0} \tilde{g}_{f,i}}\cosh 
    \left(
  \sqrt{\frac{\tilde{g}_{f,i}}{\pi ^3 }}\frac{e^{\frac{\pi ^3}{24 \tilde{g}_{f,i} \ell_0}}}{ \tilde{J}^V_i}
    \right) + \sqrt{\frac{\pi ^3 }{\tilde{g}_{f,i}}}\sinh 
    \left(
  \sqrt{\frac{\tilde{g}_{f,i}}{\pi ^3 }}\frac{e^{\frac{\pi ^3}{24 \tilde{g}_{f,i} \ell_0}}}{ \tilde{J}^V_i}
    \right).
\end{equation}
On the other hand, if $\ell_K \gg \ell_f$, the Kondo scale is given by Eq. \eqref{eq:ellKcross2} with $\tJV_i \to \tJV_i e^{-\frac{\pi ^3}{24 \tilde{g}_{f,i} \ell_0}}$:
\begin{equation}
    \begin{aligned}
    \ell_K = & 
    e^{\frac{\pi ^3}{24 \tilde{g}_{f,i} \ell_0}}\frac{2  e^{\frac{3}{4}} }{\tilde{J}^V_i}\sqrt{\frac{3\tilde{g}_{f,i} \ell_0}{\pi^3}}-\frac{3}{4} \left(\log \left(\frac{\sqrt{\frac{3\tilde{g}_{f,i}\ell_0}{\pi ^3}} e^{\frac{\pi ^3}{24 \tilde{g}_{f,i} \ell_0}}}{ 6\tilde{J}^V_i\ell_0 (\ell_0+2\sqrt{3\ell_0})}\right)+\frac{3}{4}\right)^2
    \\ & +\sqrt{3} \sqrt{\ell_0+2\sqrt{3\ell_0}} \log \left(\frac{\sqrt{\frac{3\tilde{g}_{f,i}\ell_0}{\pi ^3}} e^{\frac{\pi ^3}{24 \tilde{g}_{f,i} \ell_0}}}{ 6\tilde{J}^V_i\ell_0 (\ell_0+2\sqrt{3\ell_0})}\right) +\frac{3}{4} \sqrt{3} \sqrt{\ell_0+2\sqrt{3\ell_0}}+2\sqrt{3\ell_0}
    \end{aligned}
    \label{eq:Kondoscalefinalcase}
\end{equation}
to the leading order.

\end{document}